\documentclass[twocolumn,superscriptaddress, pre,showpacs,longbibliography]{revtex4-1}   \usepackage{amsfonts} \usepackage{amssymb} \usepackage{amsmath}
\usepackage{amsthm} \usepackage{dsfont} \usepackage{graphicx} 

\usepackage{hyperref}

  \newcommand{\be}{\begin{equation}} \newcommand{\ee}{\end{equation}}

\begin{document} \title{Slowest local operators in quantum spin chains}

\author{Hyungwon Kim} \affiliation{Physics Department, Princeton University, Princeton, NJ 08544, USA} \affiliation{Department of Physics and Astronomy, Rutgers University, Piscataway, NJ 08854, USA}

\author{Mari Carmen Ba\~{n}uls} \affiliation{Max-Planck-Institut f$ {\ddot{u}}$r Quantenoptik, Hans-Kopfermann-Str. 1, 85748 Garching, Germany}

\author{J. Ignacio Cirac} \affiliation{Max-Planck-Institut f$ {\ddot{u}}$r Quantenoptik, Hans-Kopfermann-Str. 1, 85748 Garching, Germany}

\author{Matthew B. Hastings} \affiliation{Station Q, Microsoft Research, Santa Barbara, CA 93106-6105, USA} \affiliation{Quantum Architectures and Computation Group, Microsoft Research, Redmond, WA 98052,
USA}

\author{David A. Huse} \affiliation{Physics Department, Princeton University, Princeton, NJ 08544, USA}

\begin{abstract} We numerically construct slowly relaxing local operators in a nonintegrable spin-1/2 chain. Restricting the support of the operator to $M$ consecutive spins along the chain, we
exhaustively search for the operator that minimizes the Frobenius norm of the commutator with the Hamiltonian. We first show that the Frobenius norm bounds the time scale of relaxation of the operator at
high temperatures. We find operators with significantly slower relaxation than the slowest simple ``hydrodynamic'' mode due to energy diffusion. Then, we examine some properties of the nontrivial slow
operators. Using both exhaustive search and tensor network techniques, we find similar slowly relaxing operators for a Floquet spin chain; this system is hydrodynamically ``trivial'', with no conservation
laws restricting their dynamics. We argue that such slow relaxation may be a generic feature following from locality and unitarity. \end{abstract}


\maketitle

\section{Introduction}
It has been proposed that an isolated quantum many-body system with a few local conservation laws can still thermalize, in the sense that local observables approach their thermal
equilibrium values \cite{Deutsch:1991,Srednicki:1994,Rigol:2008} set by conserved quantities. Quantum many-body systems with extensive number of local conserved quantities (integrable systems) still relax
to stationary states, which are described by the diagonal ensemble \cite{Linden:2009}. These stationary states of integrable systems can sometimes be equivalent to a generalized Gibbs ensemble
\cite{Rigol:2007,Calabrese:2011,Gogolin:2011,Fagotti:2014} but not in general \cite{Wouters:2014,Pozsgay:2014,Goldstein:2014-2}. See Ref. ~\onlinecite{Mestyan:2014} for a review. Experimental studies have
increased the interest in these questions \cite{Polkovnikov:2011, Yukalov:2011}. One proposed theoretical explanation of thermalization of a generic quantum many-body system is the eigenstate
thermalization hypothesis (ETH) \cite{Deutsch:1991,Srednicki:1994,Rigol:2008,Santos:2010,Rigol:2012,Kruczenski:2013,Beugeling:2014,Sorg:2014,Kim_ETH,Goldstein:2014}, which argues that many-body eigenstates
of thermalizing nonintegrable Hamiltonians have local reduced density operators that are thermal; then, at long times, dephasing between different energy eigenstates brings small subsystems to thermal
equilibrium.

However, not all systems show this local thermalization in an accessible time scale \cite{Banuls:2011}. One possibility is that the ETH is false for the system of Ref.~\onlinecite{Banuls:2011} but it seems
unlikely \cite{Kim_ETH}; another possibility is that the time scales required to thermalize locally are too long to be numerically accessible.  A question is: how can such slow thermalization arise, when a
nonintegrable system like the one studied has no conserved local quantities other than energy?

In this work, we illustrate how such slow relaxation can emerge by showing that indeed slow, almost-conserved local operators {\it are} present in many nonintegrable systems (note that Ref.
\onlinecite{Banuls:2011} studied slowly thermalizing {\it initial states}). We construct these operators numerically, by explicitly searching for operators with a small commutator with the Hamiltonian. For
numerical reasons discussed below, much of our work focuses on the Frobenius norm rather than the operator norm to measure the commutator, but we also discuss the operator norm and make a connection to
Ref. \onlinecite{Banuls:2011}.

One such operator with small commutator is the one that results from the thermal diffusion due to a spatially-smooth inhomogeneity in the energy density.  This operator and its commutator with the
Hamiltonian match the expectations from diffusive hydrodynamics, with the square of the Frobenius norm of the commutator decreasing as $\sim M^{-2}$ for the slowest such operator on a subregion containing
$M$ spins. However, we show numerically that these are {\it not} the operators with the smallest commutator. We construct operators whose commutator with the Hamiltonian, for the accessible system sizes of
$M \sim 100$, is quantitatively and substantially smaller (slower) than that of the simple diffusive mode, and appears to decrease with a {\it larger} power of length than the diffusive $\sim M^{-2}$. Thus
we find ``unexpected almost-conserved quantities" for these systems.

To further understand the presence of such approximately conserved quantities we turn to a Floquet spin chain where energy is not conserved. We also find slowly relaxing operators in this Floquet system.
In fact, we argue that some slowly relaxing operators must be present in any Floquet system or more generally in any quantum circuit. However, these slowly relaxing operators are in a sense morally similar
to the slowly relaxing operator in a Hamiltonian system describing energy fluctuations: these slow operators are present in any such system, so long as the unitary dynamics is local.  They themselves do
not inhibit relaxation of the local density matrices ``as fast as possible" i.e., on a time scale proportional to the length of the interval; (see Ref.~\onlinecite{Brandao:2012} for a proof that this
happens for random local circuits). Thus, the real surprise is our numerical observation that there are other operators in some nonintegrable Hamiltonian systems (and possibly in some quantum circuits)
with even slower relaxation.

The rest of this work is organized as follows. In section II, we study a model Hamiltonian system, where energy conservation is the only local conservation law. We explain how to find the local operator of
length $M$ that gives the smallest Frobenius norm of the commutator with the Hamiltonian and we connect such quantity to thermalization time of local operators. Then, we show how the Frobenius norm
decreases as we increase the length $M$ of the operator. By comparing with the simple diffusive mode, we establish that there exist some local operators that do thermalize slower than diffusion. Next, we
study the operator norm of the commutator with the Hamiltonian to make a more natural connection to the previous work. In section III, we study systems without any local conservation law. We study a
Floquet system, which is a natural counterpart to the Hamiltonian system without energy conservation, and do the same analysis to find slow operators. We find slowly relaxing operators, whose
thermalization time is bounded below by $1/M$. Then, we show that this phenomenon is generally true by finding the same scaling in random quantum circuits. In section IV, we summarize what we have found.

\section{Hamiltonian system: Presence of local conservation law}
\subsection{Model}
As a nonintegrable model Hamiltonian, we choose a spin-1/2 Ising chain with both longitudinal and transverse fields:
\begin{align}
H = \sum_{i = -\infty}^{\infty} g\sigma^x_i + h\sigma^z_i + \sigma^z_i \sigma^z_{i+1} ~,
\label{eq:Hamiltonian}
\end{align}
where $\sigma^x_i$ and $\sigma^z_i$ are Pauli matrices of the spin
at site $i$. Ref.~\onlinecite{Banuls:2011} has found a nonthermalizing state for this model within the accessible time scale. We choose $(g,h) = (0.905, 0.809)$, at which this model is known to be robustly
nonintegrable even for a relatively small system size \cite{Kim:2013}. See appendix for another set of parameters and different model Hamiltonians.

\subsection{Method}
\begin{figure}
\includegraphics[width=1.0\linewidth]{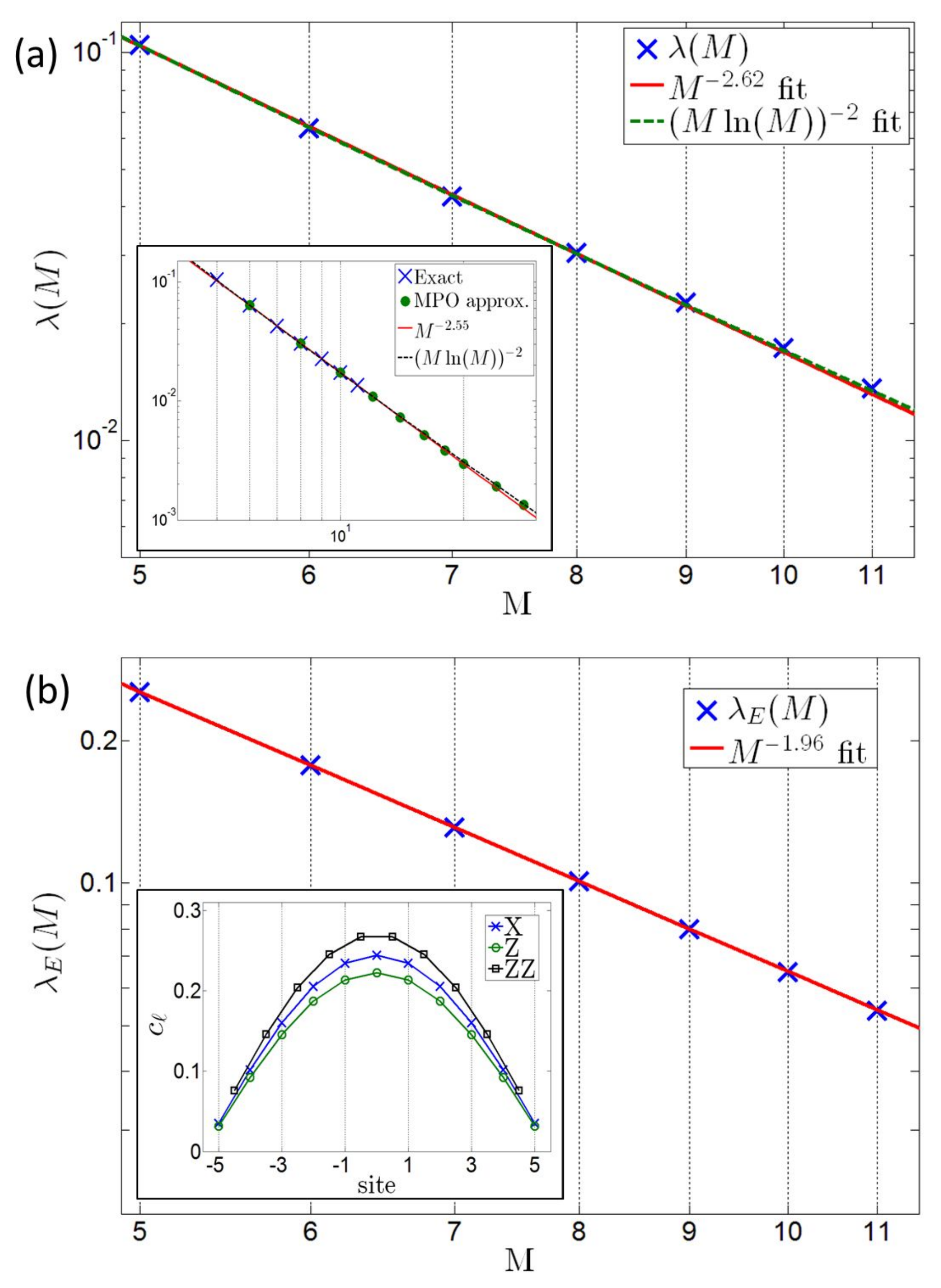}
\centering \caption{(color online) (a) Decay of $\lambda{(M)}$, minimum of Eq. \eqref{eq:minimize}, as a function of $M$ with
power-law and logarithmic correction fits.  Inset: $\lambda(M)$ vs. $M$ for best MPO results (green dots). It can also be well fitted by both power-law with almost the same exponent (-2.62 for exact and
-2.55 for MPO) and logarithmic correction. (b) $\lambda_E(M)$, the minimum of Eq. \eqref{eq:minimize} with only terms in the Hamiltonian, vs. $M$. $M^{-2}$ decay is consistent with thermal diffusion.
Inset: Structure of the optimal operator for $M = 11$. $X = \sigma^x_\ell$, $Z = \sigma^z_\ell$, and $ZZ = \sigma^z_\ell \sigma^z_{\ell+1}$. We locate the $\sigma^z_\ell \sigma^z_{i+1}$ term at $\ell+1/2$.
Coefficients are normalized as $\sum_\ell c_\ell^2 = 1$. It shows a clear sinusoidal energy modulation with the expected wavelength $2M$, and with the relative ratios equal to those in the Hamiltonian.}
\label{fig:hamiltonian}
\end{figure}

We consider local operators supported on a finite interval of $M$ consecutive sites. (We have also studied translationally invariant operators
\footnote{We have also performed similar analysis in the infinite chain for translationally invariant operators where the action of the local operator is the same on all sites. Most of our main results (Figures \ref{fig:hamiltonian} (a), \ref{fig:floquet} (a)) remain true. Since it is easier to visualize local operators and to directly compare with diffusive energy mode for translationally non-invariant operators, we present the results on an infinite chain where the local operator is placed on $M$ specific consecutive sites. See the Appendix.}.)
Since every traceless Hermitian operator $\hat{A}_M$ can be expressed by a linear combination of
$4^M - 1$ traceless Hermitian basis operators, we write
\begin{align}
\hat{A}_M = \sum_{\ell = 1}^{4^M - 1} c_\ell \hat{O}_{(M,\ell)} ~,
\label{eq:basis_expand}
\end{align}
where $c_\ell$ is a real number and $\hat{O}_{(M,\ell)}$
is the corresponding basis operator. We choose $\hat{O}_{(M,\ell)}$
to be mutually orthogonal using the Hilbert-Schmidt inner product so that $\mathrm{tr(\hat{O}_{(M,\ell)} \hat{O}_{(M,k)})} = 0$ for
$\ell\neq \mathrm{k}$.

The dynamics of an operator comes from the commutator with the Hamiltonian.
Therefore, we want to {\it minimize} the magnitude of $[\hat{A}_M, H]$ to construct a slowly relaxing local operator of length
$M$. We use the square of the Frobenius norm, $\mathrm{tr(\hat{O}\hat{O}^\dag)}$, to quantify the commutator since it gives a quadratic form  (Eq. \eqref{eq:minimize}) of which we can readily find the
minimum. Although the operator norm generally controls the dynamics of arbitrary states at arbitrary time, its numerical minimization is very challenging. As we show below, the (square of) Frobenius norm
can actually bound the thermalization time scale at infinite temperature, where we expect the dynamics to be fastest. Furthermore, we can get the upper bound of the operator norm by using the operators
that minimize the Frobenius norm. The behavior of upper bounds is consistent with the results we obtained using the Frobenius norm.

We minimize the following:
\begin{align}
\label{eq:minimize}
f(\hat{A}_M) &= \frac{\mathrm{tr([\hat{A}_M,H][\hat{A}_M,H]^\dag)}}{\mathrm{tr(\hat{A}_M\hat{A}^\dag_M)}} \nonumber\\
&= \sum_{\ell,\mathrm{k}}\frac{c_\ell c_{\mathrm{k}} \mathrm{tr([\hat{O}_{(M,\ell)},H][\hat{O}_{(M,k)},H]^\dag)}}{\sum_{\mathrm{j}} c_{\mathrm{j}} ^2 \mathrm{tr((\hat{O}_{(M,j)})^2)}} ~.
\end{align}
We define
$\lambda(M)$ to be the minimum of $f(\hat{A}_M)$: $\lambda(M) = \mathrm{min}\{f(\hat{A}_M)\}$, and we call the corresponding $\hat{A}_M$ the slowest operator acting on $M$ sites. Since the Hamiltonian has
time-reversal symmetry, we can consider even and odd operators under time-reversal separately. It turns out that for $M\geq 4$, the minimizer of $f(\hat{A}_M)$ always comes from the even sector. Up to $M =
11$, we obtain exact results and for larger $M$ we minimize Eq. \eqref{eq:minimize} using a matrix product operator (MPO) ansatz \cite{Pirvu:2010} for $\hat{A}_M$. For $M\leq 28$, we find the values of
$\lambda(M)$ have converged within 1$\%$ error.

First, let's understand the physical meaning of $f(\hat{A}_M)$. We consider an initial mixed state $\rho = I/Z + \epsilon \hat{A}_M$, where $Z$ is the normalization factor, $I$ is the identity and
$\epsilon$ is chosen to make $\rho$ nonnegative. $\hat{A}_M$ serves as a small inhomogeneity in the infinite temperature ensemble and is assumed to have unit Frobenius norm. Let's define $a_M(t)$ as the
expectation value of $\hat{A}_M/\epsilon$ at time $t$: $a_M(t) = (1/\epsilon)\mathrm{tr}(\rho \hat{A}_M(t)) = \mathrm{tr}(\hat{A}_M \hat{A}_M(t))$, where $\hat{A}_M(t)$ is in the Heisenberg picture. [Note
that $a_M(0) = 1$.] Using Cauchy-Schwarz inequality, we have the following:
\begin{align}
\left|\frac{d^2 a_M(t)}{dt^2}\right| = |\mathrm{tr}([\hat{A}_M(t),H][\hat{A}_M,H])| \leq f(\hat{A}_M) ~,
\label{eq:second_derivative}
\end{align}
Then, we can bound the distance between $a_M(t)$ and $a_M(0)$.
\begin{align}
|a_M(t) - a_M(0)| = \left|\int^t_0 d\tau \int^\tau_0 d\tau' \frac{d^2 a_M(\tau')}{d\tau'^2}\right| \leq \frac{f(\hat{A}_M) t^2}{2} ~.
\end{align}
In thermodynamic limit and finite $M$, the thermal expectation value is $a_M^{th} = \rm{tr}(\hat{A}_M) = 0$. Therefore, the
following inequality holds:
\begin{align}
|a_M(t) - a_M^{th}| &\geq |a_M(0) - a_M^{th}| - |a_M(0) - a_M(t)|  \nonumber\\
& \geq 1 - \frac{f(\hat{A}_M)t^2}{2} ~.
\label{eq:hamiltonian_timescale}
\end{align}
Consequently, $f(\hat{A}_M)$ bounds the thermalization time scale $\tau$ of $\hat{A}_M$ from below by $\tau \geq f(\hat{A}_M)^{-1/2}$, and small $\lambda(M)$ implies a long thermalization time of
$\hat{A}_M$. In addition, since $\lambda(M)$ is the minimum of Eq. \eqref{eq:second_derivative} at $t =0$, the optimal $\hat{A}_M$ is {\it the} slowest operator at early time.

\subsection{Results and Comparison with energy diffusion}
Figure \ref{fig:hamiltonian} (a) plots $\lambda (M)$. It is clear that $\lambda(M)$ decreases with $M$ as it should. The data can be well-fitted by
two functional forms; power-law decay with exponent $2.62$ and a logarithmic correction to $1/M^2$, which is, more precisely,
a fit to $a/(M\ln(bM))^2$, with $a$ and $b$ as fitting parameters. In either
case, the rate of decrease with $M$ is {\it faster} than $1/M^2$, which is the scaling of the slowest diffusive energy mode as we show now.

A slowly relaxing energy mode can be constructed by considering an energy modulation of wavelength $2M$:
\begin{align}
\hat{E}_M = \sum_i c_M(i) h_i &= \sum_{i=-M/2}^{M/2} \cos\left(\frac{i\pi}{M}\right)(g \sigma^x_i + h\sigma^z_i)\nonumber\\
&\quad+ \sum_{i=-M/2}^{M/2-1} \cos\left(\frac{(i+1/2)\pi}{M}\right)\sigma^z_i\sigma^z_{i+1} ~,
\label{eq:energy_modulation}
\end{align}
where $h_i$ is the energy
density operator ($g \sigma^x_i + h\sigma^z_i + 1/2(\sigma^z_{i-1}\sigma^z_i + \sigma^z_i\sigma^z_{i+1}$)) and $c_M(i)$ is the cosine modulation function restricted to lie in $[-M/2,M/2]$. Since the energy
is conserved, we can use the continuity equation: $d h_i/dt = -\nabla \cdot {\bf j}_i$, where ${\bf j}_i$ is the energy current density at site $i$:
${\bf j}_i = g(\sigma^y_i \sigma^z_{i+1} - \sigma^y_{i+1} \sigma^z_i)$.
Combining this with the Heisenberg equation of motion, we have the following.
\begin{align}
 i[H,\hat{E}_M] &= \frac{d}{dt} \hat{E}_M = -\sum_i c_M(i) \partial_i \cdot {\bf j}_i \\
 &= \sum_i {\bf j}_i \partial_i c_M(i) \simeq -\frac{\pi}{M} \sum_i s_M(i) h_i,
\end{align}
where $\partial_i$ is the discrete spatial derivative and $s_M(i)$ is the sine modulation. Therefore,
\begin{align}
\frac{\mathrm{tr}([H,\hat{E}_M][H,\hat{E}_M]^\dag)}{\mathrm{tr}(\hat{E}_M \hat{E}_M^\dag)} \sim 1/M^2 ~.
\end{align}

Here, we adapt a more conservative approach. We do the same numerical search as before but restrict the operator space within the terms in the Hamiltonian so we only consider $3M-1$ basis operators instead
of $4^M-1$. Figure \ref{fig:hamiltonian} (b) plots $\lambda_E(M)$, the minimum of $f(\hat{A}_M)$ in this restricted space. As expected, we have almost perfect $1/M^2$ scaling. Since there are only a few
basis operators, we can easily look at the details of the structure of the optimal operator. The inset of Figure \ref{fig:hamiltonian} (b) is the structure of the optimal operator in terms of the local
terms in Hamiltonian. It is indeed of the form of Eq.~\eqref{eq:energy_modulation} and thus the energy modulation of the longest wavelength is the slowest energy mode. Note also the fact that apart from
displaying a different scaling, $\lambda(M)$ is much smaller than $\lambda_E(M)$.

This hydrodynamic description, however, only considers operators that are {\it linear} in energy density operators. Combinations of higher power of energy density terms, such as $h_i h_j$, $h_i h_j h_k$,
$\ldots$, may yield the slowest operator given by exhaustive search. Although we are unable to obtain the exact slowest operators, we find a deviation of the scaling of $\lambda(M)$ from $M^{-2}$ by adding
nonlinear energy density operators. Furthermore, most of the weight (measured by Hilbert-Schmidt norm) of the slowest operator turns out to be energy density modulation. In actuality, we can obtain slow
operators (not slowest) by ``dressing'' the modulated form. See Appendix for details.

\subsection{Operator Norm}

The operator norm is mathematically more convenient for studies of time scale since we can directly interpret the relaxation of the operator in terms of Lieb-Robinson bounds \cite{Lieb:1972,Bravyi:2006}.
Optimizing the operator norm numerically is, however, very challenging. Therefore, we have minimized the (square of) Frobenius norm. Nevertheless, once we have an operator, it is easy to compute the
operator norm of the commutator with the Hamiltonian.

Our first task is to relate the operator norm with the thermalization time scale. Let us assume that $\hat{A}_M$ satisfies the following:
\begin{align}
||[\hat{A}_M, H]|| \leq \chi(M),
\end{align}
where
$H$ is the Hamiltonian and $||\ldots||$ means the operator norm of the argument and $\chi(M)$ is some nonnegative valued function. Then, using the Heisenberg equation of motion, we have
\begin{align}
\bigg|\bigg|\frac{d}{dt} \hat{A}_M(t)\bigg|\bigg| &= \bigg|\bigg| e^{-i H t} \left(\frac{d}{dt} \hat{A}_M (t)\right) e^{i H t} \bigg|\bigg| \nonumber\\
&= ||[\hat{A}_M(t = 0), H]|| \leq \chi(M) \\
||\hat{A}_M(t) - \hat{A}_M(0)|| &= \bigg|\bigg|\int_0^t \left(\frac{d}{d\tau} \hat{A}_M(\tau)\right)d\tau \bigg|\bigg| \nonumber\\
&\leq \int^t_0 \big|\big|\frac{d}{d\tau} \hat{A}_M(\tau)\big|\big|d\tau \leq \chi(M) t ~,
\end{align}
where we have used the fact that $e^{-iHt}$ is a norm-preserving unitary operator.
This inequality bounds the distance of an operator evolving under Hamiltonian dynamics from
its initial configuration. Note that we have previously used the {\it square} of the Frobenius norm to bound the distance from the thermal value,
where we find the bound of the thermalization time is given
by $\sqrt{\lambda(M)}$.

Next, we consider an initial state $\rho$ such that
\begin{align}
|\langle \hat{A}_M \rangle_0 - \langle \hat{A}_M \rangle_\beta| = \gamma(M) ~,
\label{eq:thermal_distance}
\end{align}
where $\langle \ldots \rangle_0$ is the expectation value of the initial condition,
$\langle \ldots \rangle_\beta$ is the thermal expectation value and $\gamma(M)$ is some nonnegative valued function.
 Now we can estimate
the distance between the thermal expectation value and the expectation value at time $t$ (for $t \leq \gamma(M)/\chi(M)$):
\begin{align}
&|\langle \hat{A}_M \rangle_t - \langle \hat{A}_M \rangle_\beta| \nonumber\\
&\geq \left | |\langle \hat{A}_M\rangle_0 - \langle \hat{A}_M\rangle_\beta | -|\langle \hat{A}_M\rangle_t - \langle \hat{A}_M\rangle_0 |\right | \nonumber\\
&\geq \gamma(M) - t \chi(M)
\end{align}
where $\langle \ldots \rangle_t$ is the expectation value at time $t$. Therefore, if we have a sequence of $M$-body operators $\{ \hat{A}_M \}$ for which the operator norm of the commutator
with the Hamiltonian decays fast with $M$ and an initial state which does not allow a fast decrease of $\gamma(M)$, the time scale of thermalization of $\hat{A}_M$ is
\begin{align}
\tau_M \geq \frac{\gamma(M)}{\chi(M)} ~.
\label{eq:op_norm_time}
\end{align}
In particular, if $\chi(M)$ decreases faster than a power law with $M$, thermalization may take longer than polynomial time in contrast to
the case of diffusion.

\begin{figure}
\includegraphics[width=1.0\linewidth]{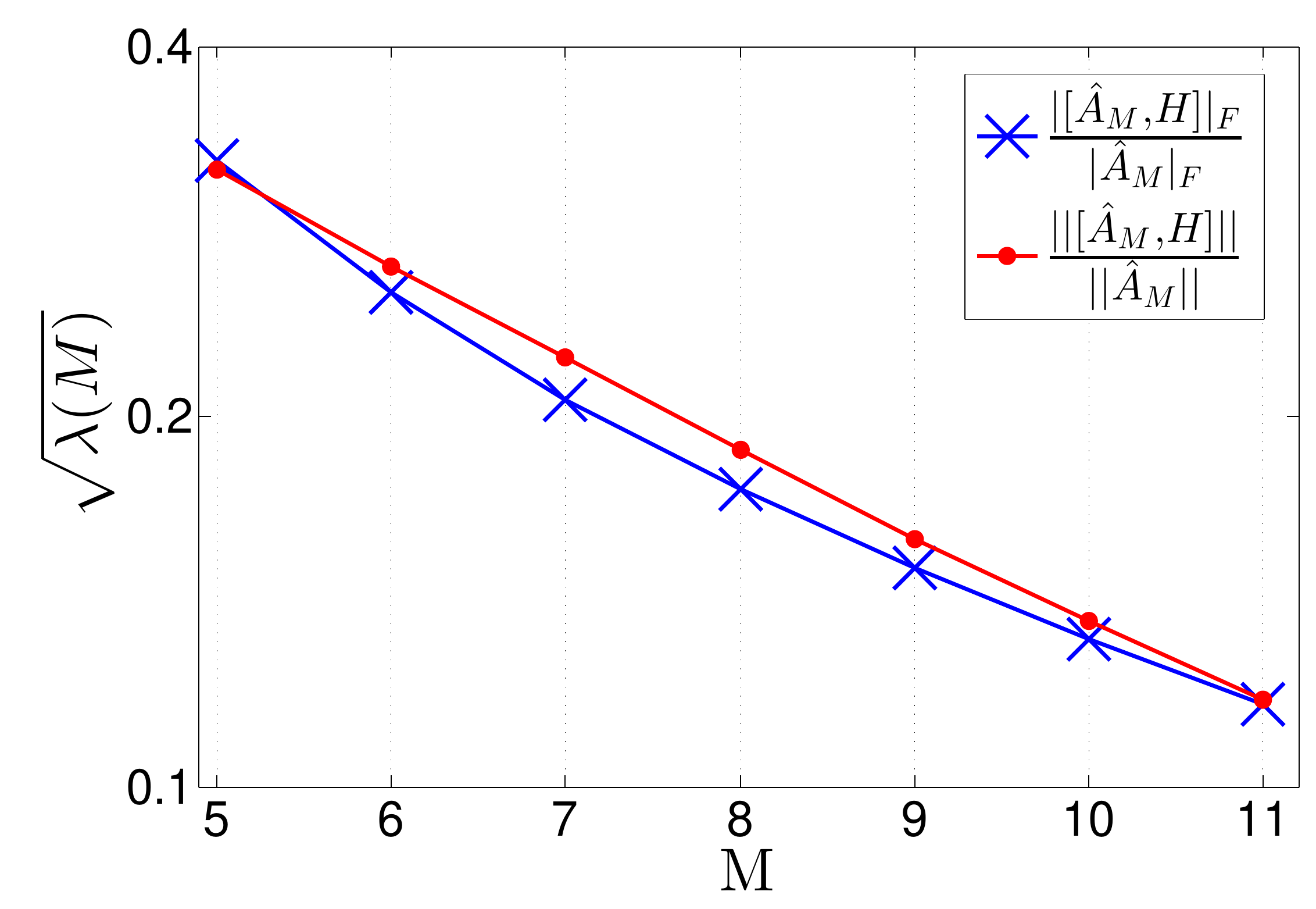}
\centering \caption{(color online) Normalized
magnitude of $[\hat{A}_M, H]$ measured by the Frobenius norm ($|\ldots|_F$) and the operator norm ($||\ldots||$) in log-linear plot. $\hat{A}_M$ is obtained by minimizing the Frobenius norm of the
commutator with Hamiltonian as is in the main text. Two norms give quantitatively similar values. }
\label{fig:op_norm}
\end{figure}

Figure \ref{fig:op_norm} plots the decay of the Frobenius norm and the operator norm of $[\hat{A}_M,H]$, where $\hat{A}_M$ is obtained in the main text by minimizing the Frobenius norm. It shows that the
value of $\lambda$ measured by the operator norm is numerically similar to that of the Frobenius norm. It appears that the operator norms computed for these operators exhibit exponential-like decrease with
$M$ but the range of $M$ is not enough to draw a conclusion.

\subsubsection{Initial State and Thermalization Time Scale}
\begin{figure}
\includegraphics[width=1.0\linewidth]{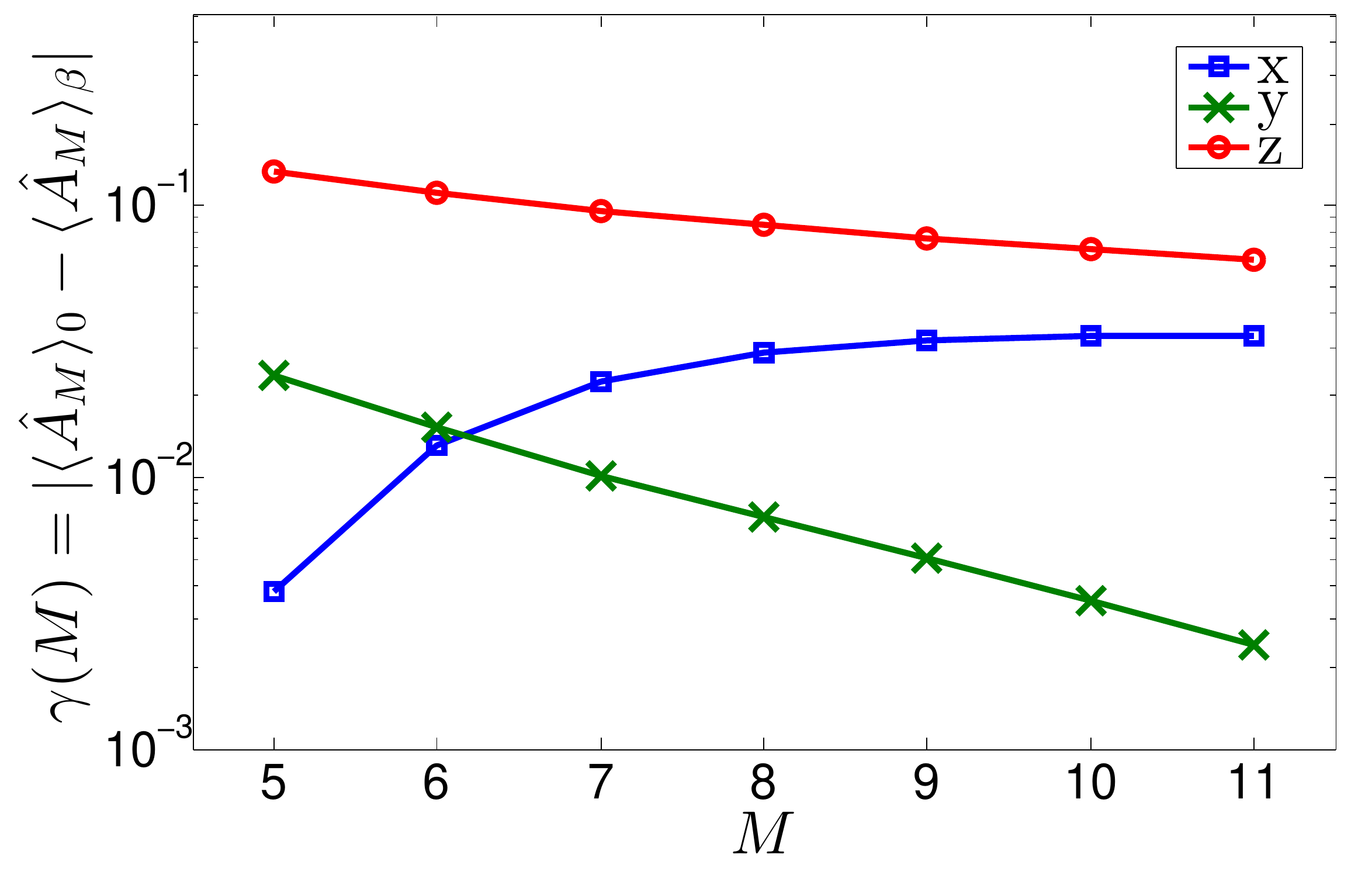}
\centering \caption{(color online) Distance between expectation
values of initial spin-polarized states and corresponding thermal states. For the $x$-polarized state (blue square), the distance does not decrease as $M$ increases. $\hat{A}_M$ is obtained by minimizing
the Frobenius norm of the commutator with Hamiltonian. Label means that direction along which the initial state is polarized. }
\label{fig:op_norm_result}
\end{figure}

Ref. \onlinecite{Banuls:2011} reports nonthermalization of spin polarized initial state along $x$ direction and weak thermalization of spin polarized initial state along $z$ direction. Having slowest
operators from the Frobenius norm, we study their expectation values with respect to spin-polarized states. Figure \ref{fig:op_norm_result} plots the distance between the expectation values of
spin-polarized initial states and corresponding thermal states. The operator $\hat{A}_M$ is obtained by minimizing the Frobenius norm as we explained before. It is noteworthy that the distance does not
decrease for $x$-polarized state. From Eq.\eqref{eq:op_norm_time} and Figs. \ref{fig:op_norm} and \ref{fig:op_norm_result}, we see that thermalization time scale increases fast for $x$-polarized state,
which is consistent with Ref.\onlinecite{Banuls:2011}. However, it needs some care to make direct connection to Ref. \onlinecite{Banuls:2011} since our operator is {\it not} translation invariant and we do
not minimize the operator norm. We leave this problem for later studies.

\section{Systems without local conservation law}
\subsection{Floquet system}
Refs.~\onlinecite{Dalessio:2014, Lazarides:2014, Ponte:2014} test whether the energy conservation is important in the slow
relaxation of local operators. We adopt the same Floquet operator used in Ref.~\onlinecite{Kim_ETH}:
\begin{align}
U_F = \exp(-i H_x \tau) \exp(-i H_z \tau) ~,
\label{eq:floquet_op}
\end{align}
where $H_x$
is the $\sigma^x$ part ($\sum_i g \sigma^x_i$) and $H_z$ is the $\sigma^z$ part ($\sum_i h \sigma^z_i +\sigma^z_i \sigma^z_{i+1}$) of the Hamiltonian.
We choose $\tau = 0.8$. Although $U_F$ does not
conserve energy, it acts {\it locally} on the system. This type of Floquet operator is shown to thermalize a local operator to the infinite temperature ensemble \cite{Kim_ETH,Prosen:2002}.

\subsection{Method}
We minimize the square of the Frobenius norm of the commutator with the Floquet operator. Up to $M = 11$, we obtain exact results and for larger $M \leq 16$ we use the MPO ansatz to
find the minimum of the following
\begin{align}
\label{eq:floquet_minimize}
g(\hat{A}_M) = \frac{\mathrm{tr}([\hat{A}_M,U_F][\hat{A}_M,U_F]^\dag)}{\mathrm{tr}(\hat{A}_M\hat{A}_M^\dag)} ~.
\end{align}
Again, we define $\lambda(M) = \mathrm{min}\{g(\hat{A}_M)\}$ and call the corresponding $\hat{A}_M$ the slowest operator.

Let's relate $g(\hat{A}_M) $ with the thermalization time scale. As in the case of Hamiltonian system, we consider the same initial state; $\rho = I/Z + \epsilon \hat{A}_M$. Since the time step in the
Floquet system is discrete, we define $a_M^{(N)} = \mathrm{tr}(\hat{A}_M(N)\hat{A}_M )$, with $a_M^{(0)} = 1$ and $\hat{A}_M(N) = U_F^{\dag N}\hat{A}_M U_F^N$. Using Cauchy-Schwarz inequality, we can show
that
\begin{align}
|a_M^{(n+1)} - a_M^{(n)}| = |\mathrm{tr}([\hat{A}_M(n),U_F]\hat{A}_M U_F^{\dag})| \leq \sqrt{g(\hat{A}_M) } ~.
\end{align}
Then, the following inequality follows:
\begin{align}
|a_M^{(N)} - a_M^{0}| \leq \sum_{n = 0}^{N-1}|a_M^{(n+1)} - a_M^{(n)}| \leq N \sqrt{g(\hat{A}_M) } ~.
\end{align}
Since the thermal expectation value $\hat{a}_M$ is indeed zero in the Floquet system
\cite{Kim_ETH}, we have,
\begin{align}
|a_M^{(N)} - a_M^{th}| \geq |a_M^{(0)}| - |a_M^{(N)} - a_M^{(0)}| \geq 1 - N \sqrt{g(\hat{A}_M) } ~.
\label{eq:floquet_timescale}
\end{align}
Therefore,
$g(\hat{A}_M)$ bounds the thermalization time of $\hat{A}_M$ from below by $N \geq g(\hat{A}_M) ^{-1/2}$.
Again, small $\lambda(M)$ implies a long thermalization time.

\begin{figure}
\includegraphics[width=1.0\linewidth]{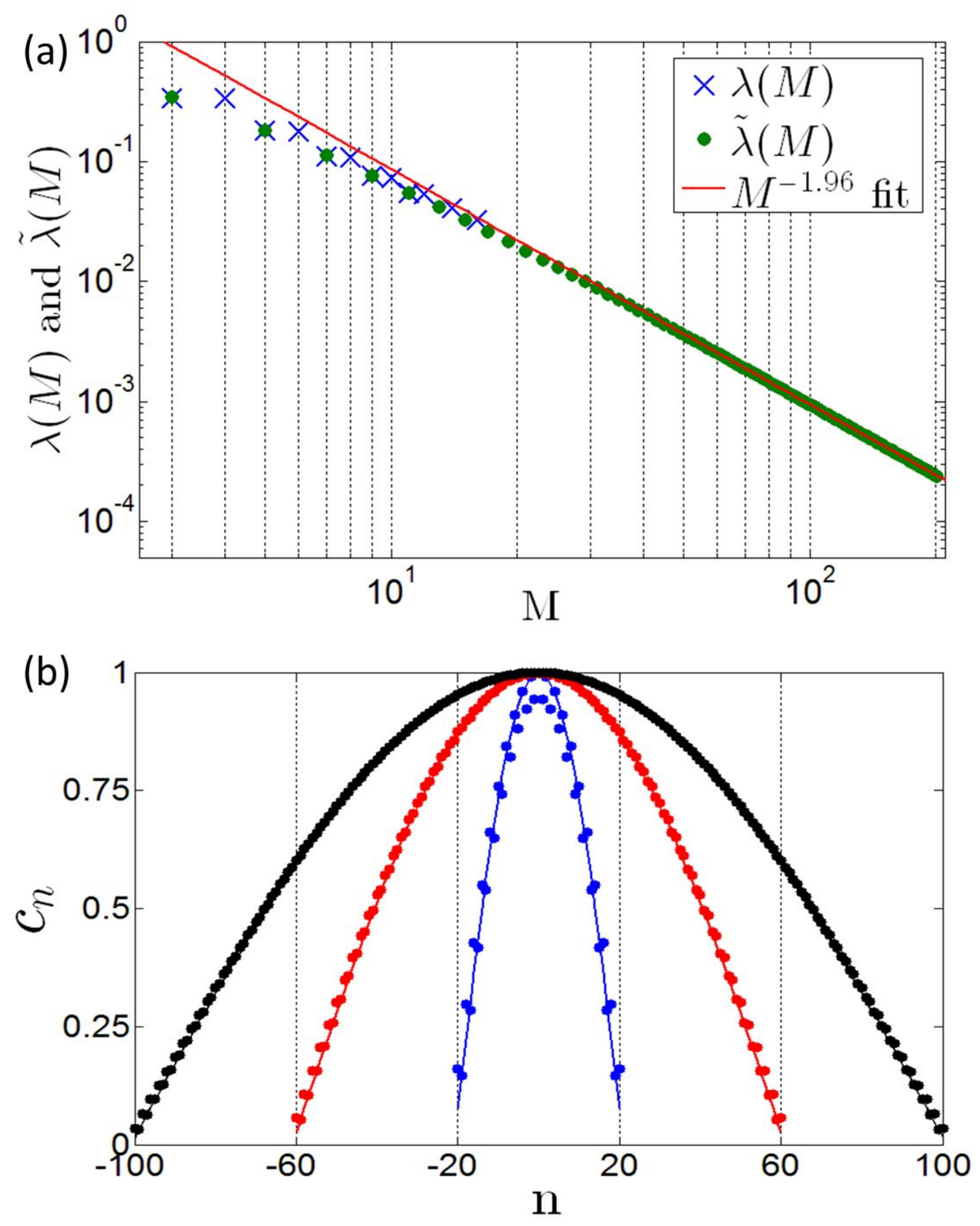}
\centering \caption{(color online) (a) Decay of $\lambda(M)$ (blue cross), the exact minimum of Eq. \eqref{eq:floquet_minimize} and
$\tilde{\lambda}(M)$ (green dot), a variational upper bound found in the form of Eq. \eqref{eq:filter}. The fractional difference between $\lambda(M)$ and $\tilde{\lambda}(M)$ is less than 1/200 for $M
\leq 11$. $\tilde{\lambda}(M)$ decreases close to $M^{-2}$ asymptotically. Note that there is a strong parity effect in the exact result: $\lambda(2N-1) \simeq \lambda(2N)$. (b) Structure of coefficients
$c_n$ (dots) for $N$ = 20 (blue, most narrow), 60 (red, intermediate width), and 100 (black, widest). A cosine function with wavelength $4N+2$ fits well except near $n = 0$ of $N = 20$.}
\label{fig:floquet}
\end{figure}

To extend to larger systems, we optimize the commutator over operators of a specific form. We consider operators in the space spanned by $U^n \hat{O} U^{- n}$ for $\hat{O}$ a traceless Hermitian operator
acting on a single site and taking a finite number of powers of unitary operator $U$ (this case, the Floquet $U_F$). Considering the operator of the following filtered form,
\begin{align}
 \tilde{A}_N=\sum_{n=-N}^N c_n U^n \hat{O} U^{- n} ~,
 \label{eq:filter}
\end{align}
where $\tilde{A}_N$ is supported on $M = 2N+1$ sites for the Floquet system, we find $\tilde{A}_N$ that gives $\tilde{\lambda}(M) =
\mathrm{min}\{\mathrm{tr}([\tilde{A}_N,U][\tilde{A}_N,U]^\dag)/\mathrm{tr}(\tilde{A}_N\tilde{A}_N^\dag)\}$ by varying the coefficients $\{c_n\}$ and $\hat{O}$. This method can be considered as a version of
the Lanczos method as it also works in a Krylov subspace, but here we use a tensor network method to approximate ${\rm tr}(\hat{O}^\dagger U^n \hat{O} U^{-n})$. Thus $\tilde{\lambda}$ is a variational
upper bound of the true minimum $\lambda$. Interestingly, this simple model with only 2$N$ + 4 real variables ($2N+1$ for $c_n$'s and 3 for $\hat{O}$) agrees very well with the available exact results ($M
=11$ or $N = 5$). Moreover, we find that the filtered $\hat{A}_N$ that gives $\tilde{\lambda}(M)$ is obtained from the same single site operator $\hat{O}$ for all values of $N$. For the given model and the
parameters, $\hat{O} = 0.04\sigma^x  - 0.66\sigma^y + 0.75\sigma^z$. Note that $\tilde{A}_N$ can only be supported on odd number of sites. It turns out that the exact operator of even support can be
approximated by a simple symmetric extension: $\tilde{A}_N\otimes \sigma^0 + \sigma^0\otimes\tilde{A}_N$, where $\sigma^0$ is an identity. Therefore, $\lambda(2N-1) \simeq \lambda(2N)$. It is noteworthy
that the filtering method is unable to capture the slowest dynamics in the Hamiltonian system.

\subsection{Results}
Figure \ref{fig:floquet} (a) is the plot of $\lambda (M)$, the minimum value of Eq. \ref{eq:floquet_minimize}, and $\tilde{\lambda}(M)$, which was obtained by the form of Eq.
\eqref{eq:filter}. $\tilde{\lambda}(M)$ asymptotically decreases very close to $M^{-2}$ and its optimal distribution of $\{ c_n\}$ is close to a cosine modulation (Figure \ref{fig:floquet} (b)), which we
expect to be a generic feature of quantum circuits with local dynamics (see the Appendix).

Comparing Figure \ref{fig:floquet} (a) with Figure \ref{fig:hamiltonian} (a), we see that $\lambda(M)$ decreases {\it slower} than that in Hamiltonian cases, which indicates that the Floquet system
thermalizes the local operator {\it faster}. Since the only apparent difference between the Hamiltonian system and the Floquet system is the existence of the energy conservation, we attribute this faster
relaxation to the absence of conservation law. Nevertheless, Figure \ref{fig:floquet} clearly exhibits that the rate of relaxation in the Floquet system becomes slower as the support $M$ increases, thus
$\hat{A}_M$ is again an approximately conserved quantity.

\subsection{Quantum Circuit}
\begin{figure}
\includegraphics[width=1.0\linewidth]{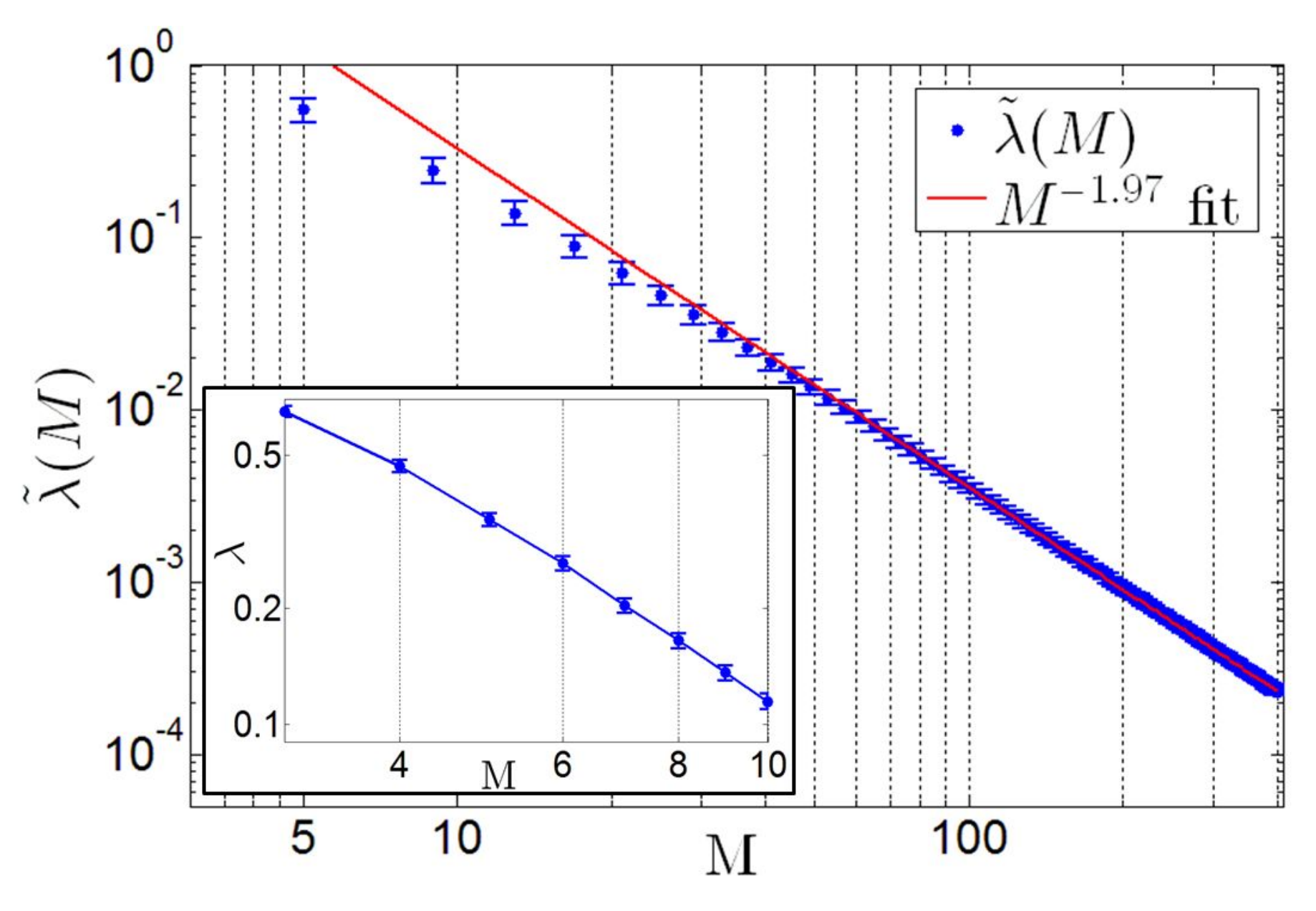}
\centering \caption{ (color online)  $\tilde{\lambda}(M)$ for a random circuit model computed by
matrix product method with the filtered of operators (Eq. \eqref{eq:filter}). $1/2\sigma^z$ is chosen for the single site Hermitian operator. We averaged 50 realizations of random circuits. Asymptotically,
the data follows very closely $M^{-2}$. Inset: $\lambda(M)$ is the result of exhaustive search without using the filtered form for the same model. These values cannot be matched by the simple filtered
operator we consider.}
\label{fig:random_circuit}
\end{figure}

To determine whether this phenomenon is more generally true, we also study a family of quantum circuits, each composed of two rounds, where in the first round, gates act on pairs of sites
$...,(1,2),(3,4),...$ and on the second round gates act on pairs $...,(0,1),(2,3),(4,5),...$ We choose all gates in a given round to be the same, but choose them randomly. The results are shown in
Fig.~\ref{fig:random_circuit}. We find again that even in this random case, slow operators are present.

To study larger systems, we again apply the matrix product method to the same form of the operators:
\begin{align}
\tilde{A}_N=\sum_{n=-N}^N c_n U^n \hat{O} U^{- n} ~.
\label{eq:filter}
\end{align} Since
our random circuit consists of two alternating non-commuting unitaries, the support of $\tilde{A}_N$ is now $4N+1$ instead of $2N+1$. In this random circuits, generally there is no single site Hermitian
$\hat{O}$ that matches the exact calculation for a small system size, unlike in the Floquet operator we studied in the main text. Therefore, we just choose $\hat{O} = 1/2\sigma^z$ as a single site
Hermitian operator. The weights $c_n$ again agree very well a cosine shape and $\lambda$ decays close to $1/M^2$. This hints that $1/M^2$ would be a generic feature.

Such a decay $1/M^2$ would be exact with a cosine if ${\rm tr}(\hat{O} U^n \hat{O} U^{- n})=0$ for all $n \neq 0$. We are unable to prove this decay in general but show weaker results in appendix. Finally,
we do not preclude possibilities that there might be some quantum systems without local conservation law where $\lambda(M)$ decreases faster than $1/M^2$, which would be very interesting.

\section{Summary and Outlook}
We have numerically constructed a series of local operators that relax slower than local energy fluctuations do. Although we can approximate these slow operators by adding
nonlinear energy density operators or assuming special filtered forms, their exact origin requires further exploration. We have also performed the same analysis in similar systems without energy
conservation, e.g. Floquet systems, again finding slowly relaxing operators. These operators present a new class of observables; if they can be studied experimentally, they may reveal unsuspected slow
relaxation.

Our method is by no means restricted to the particular Hamiltonians or Floquet operators we studied or to non-translationally invariant operators. As explained in the Appendix, it is very easy to apply
this method to find instead the slowest translationally-invariant operator. For any given any spin Hamiltonian, Floquet operator, or quantum circuit, our method should find the slowest local operator of
given length $M$. If unknown local conserved operators in terms of spin operators exist, this procedure should be able to detect them. Therefore, this may be used to find out unknown local conserved
operators or unsuspected slow dynamics (e.g. quasi many-body localization in translation invariant systems \cite{Roeck:2014,Yao:2014}) if the system of interest has such things.

\section{Acknowledgement}
We thank Fabian Essler and Ehud Altman for stimulating discussions and Stefan K\"{u}hn for help in numerical algorithms. HK is grateful for the support and hospitality of
Max-Planck-Institute f\"{u}r Quantenoptik, where this work has begun and Korea Institute for Advanced Study, where part of this manuscript was written. HK is funded by NSF DMR- 1308141. MCB and JIC were
partially funded by the EU through the SIQS integrated project.

\appendix
\section{Structure of Slowest Operators and Construction of slow operators in Hamiltonian System}
The main conclusion of this work is that the minimum value of the commutator with Hamiltonian
measured by Frobenius norm decreases faster than expected from hydrodynamics type arguments as we increase the range of operators $M$. First, let's understand why this is nontrivial.

There are exponentially many ($4^M-1$) linearly orthogonal operators for a given M.
Thus, it may not seem surprising that there exists a sequence of operators whose commutator with the Hamiltonian show
Frobenius norm with fast decreasing scaling.
However, what is important is that $\lambda(M)$ is the {\it minimum} value of commutator with the Hamiltonian,
which is not a random sequence. For instance, it is
possible to construct a series of operators of range up to $M$ (range 1, range 2, $\ldots$, range $M$ operators),
such that their commutators with the Hamiltonian scale with a larger exponent than the one we
found.
However, they are not the operators that minimize the commutator with Hamiltonian and thus not the slowest operator at a given range. We do not call them slow operators. What we found, instead, is
that the scaling of the {\it slowest} operators is different from hydrodynamics, which is usually considered to be the slowest mode of a given range (wavelength).

\subsection{Structure of slowest operator}
Although our understanding of the nature of the slowest operators in a Hamiltonian system is incomplete, we can extract some useful information by analyzing the
operators found. First, we look at how different the slowest operator is from the energy density modulation, which is expected to be the slowest mode from hydrodynamics. Once we find the slowest operator,
we can decompose the operator in an operator basis (Eq. (2) in the main text) and study the magnitude of the different components,
$c_\ell = \mathrm{tr}(\hat{A}_M \hat{O}_{M,\ell}^\dag)/\mathrm{tr}(\hat{O}_{M,\ell} \hat{O}_{M,\ell}^\dag)$, where normalization is $\sum_{\ell=1}^{4^M-1} (c_\ell)^2 = 1$.
It turns out that the slowest operator consists mainly of linear
energy density operators; $(\sigma^x_{i}, \sigma^z_{i}, \sigma^z_{i}\sigma^z_{i+1})$. For example, when $M = 11$,
the square sum of $c_{\ell}$'s of energy density operators is 0.87, which is remarkably
large given the fact that there are only 32 such operators out of $4^{11} - 1$ basis operators.
However, the relative magnitude of $c_{\ell}$'s does not exactly follow a cosine modulation, although it
shows some similarity,  so there exists some kind of ``dressing'' to energy modulation.

\begin{figure}
\includegraphics[width=1.0\linewidth]{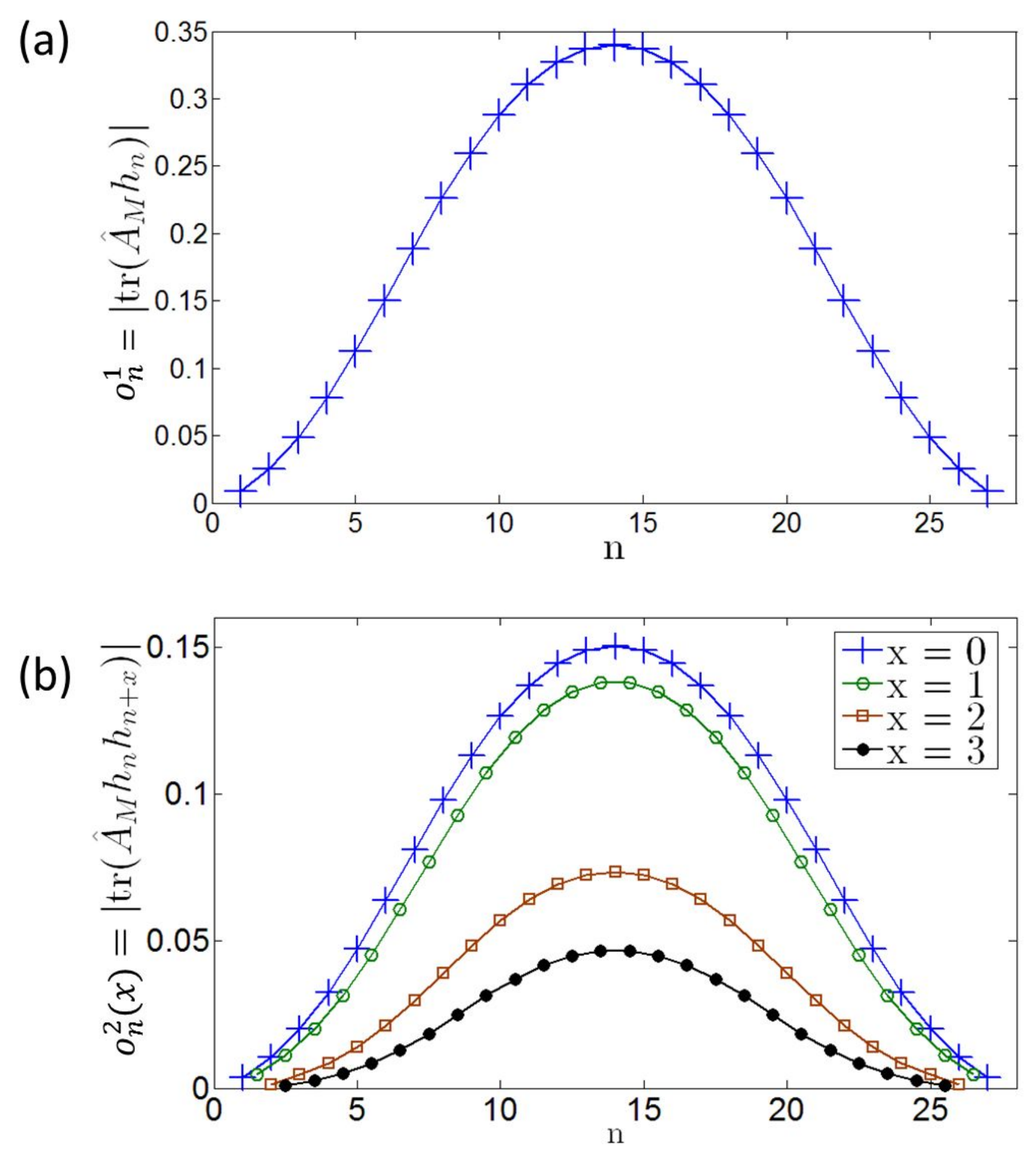}
\centering \caption{(color online) $\hat{A}_M$ is computed using MPO with bond dimension 140 and $M = 28$. (a) $o^1_n$ is the
overlap between the slowest operator $\hat{A}_M$ and the energy density operator $h_n$. The shape resembles a cosine modulation (highest in the middle and the lowest at the edges) but does not exactly
match it. (b) $o^2_n(x)$ is the overlap between the slowest operator and $h_n h_{n+x}$. The weight decreases as the separation $x$ increases. For each $x$, the shape is similar to the other cases. }
\label{fig:h_decomp}
\end{figure}

Next, we look at the overlap of the slowest operator with the energy density operators of higher powers.
It is easier to analyze the operator in terms of two site energy density operator,
$h_n$:
\begin{align}
h_n = \frac{1}{2}\left(g (\sigma^x_n + \sigma^x_{n+1}) + h(\sigma^z_n + \sigma^z_{n+1})\right) + \sigma^z_{n} \sigma^z_{n+1} ~.
\end{align}
Normalizing $h_n$ to have a unit Frobenius norm, we
compute the overlap $o_n$ (we omit the range index $M$ for brevity) resolved by the position $n$.
\begin{align}
o^1_n = |\mathrm{tr}(\hat{A}_M h_n)| ~,
\end{align}
where the superscript 1 means the overlap
of the linear order of $h_n$.
Note that $\mathrm{tr}(h_n h_{n+1}) \neq 0$ thus $o_n$ contains some contribution from $h_{n-1}$ and $h_{n+1}$. The overlap of quadratic order of energy density operator
$o^2_{n}(x)$ (superscript 2 means the quadratic order) is obtained by,
\begin{align}
o^2_{n}(x) = |\mathrm{tr} (\hat{A}_M h_n h_{n+x})| ~.
\end{align}
We can continue this to arbitrary powers of $h_n$'s.
Since $(h_n)^2$ is not identity, the $o^2_n(x)$ terms may contain some contribution already counted in $o^1_n$.
Despite this certain degree of  double-counting, this decomposition reveals another
structure in the slowest operator. Figure \ref{fig:h_decomp} is the plot of $o^1_n$ and $o^2_n(x)$ ($x = 0,1,2,3$) of the slowest operator obtained by the MPO ansatz for $M = 28$. It is clear that the
slowest operator takes more contribution from local operators (for quadratic order, smaller $x$) than non-local operators (large $x$). For the same order and $x$, their relative magnitudes are similar to
cosine modulation but not exactly the same. These features suggest that the slowest operator can be built by adding nontrivial dressing to energy density modulation and its higher powers.

In the following subsection, we build a sequence of slow operators that shows a fast decaying scaling based on the above observation. In both cases, we are able to
construct slow operators that are close to the slowest ones but unable to make them as close to have the same scaling.

\subsection{Construction of Slow Operators}

\subsubsection{Nonlinear energy density operators}
Figure \ref{fig:h_decomp} implies that the slowest operator contains nontrivial contributions from higher powers of the energy density operators.
Therefore, we build operators consisting of nonlinear powers of energy density operators. For a given power $\alpha ( = 1, 2, \ldots)$ and range $M$, we construct operators of the following form:
\begin{align}
\hat{B}_{M,1} &= \sum_n c_n h_n \\ \hat{B}_{M,2} &= \sum_{n\geq m} c_{n,m} P(n,m) h_n h_m \\
\hat{B}_{M,3} &= \sum_{n\geq m\geq l} c_{n,m,l} P(n,m,l) h_n h_m h_l ~,
\label{eq:power_H}
\end{align}
and so on. Here $P(\ldots)$ is the symmetric permutation operator that makes each term Hermitian. Here we allow $h_n$ to be identity so that $\hat{B}_{M,\alpha}$ includes all
$\hat{B}_{M,\beta}$ where $\beta \leq \alpha$. The range of each term in the summation is restricted within $M$. We optimize the coefficients $c_{\ldots}$ to have the minimum commutator with the
Hamiltonian in the Frobenius measure. Note that $\hat{B}_{M,1}$ should just be an energy modulation as we saw in the main text.

\begin{figure}
\includegraphics[width=1.0\linewidth]{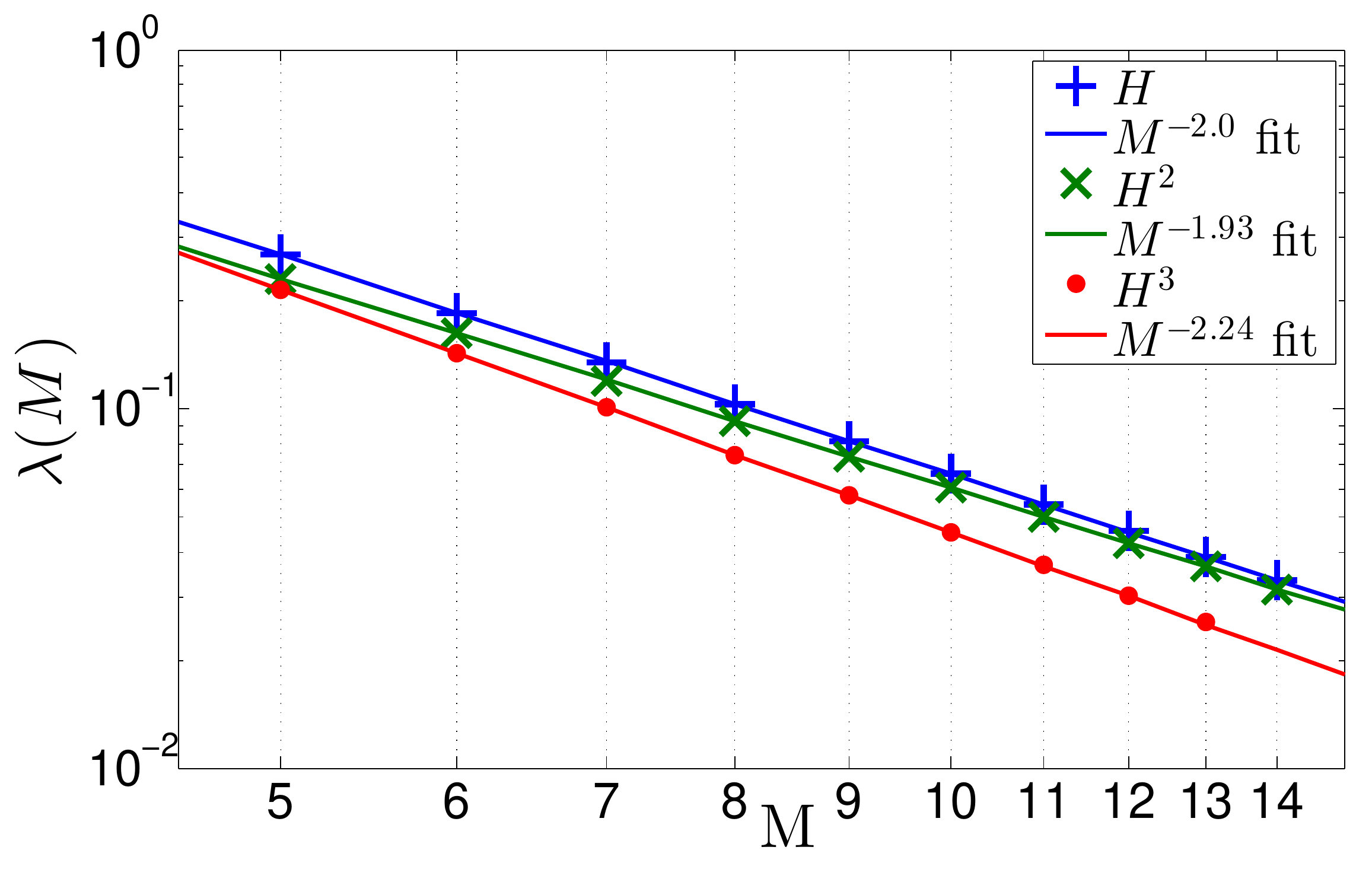}
\centering \caption{(color online) $\lambda(M)$ is the minimum value of the commutator with the Hamiltonian measured by the square of
Frobenius norm. Operators are constructed as Eq.~\eqref{eq:power_H}. For power 1, $\lambda(M) \sim M^{-2}$ as expected. For power 2, we do not make slow operators. From power 3, we start seeing a sequence
of slow operators whose scaling of $\lambda(M)$ decreases faster than $M^{-2}$.   }
\label{fig:H_power}
\end{figure}

Figure \ref{fig:H_power} is the plot of $\lambda(M)$, the minimum value of the square of the Frobenius norm of the commutator with the Hamiltonian for powers 1, 2, and 3. At linear power, $\lambda(M)$
decreases with $M^{-2}$ as expected from conventional hydrodynamics. For the quadratic order, however, we do not see a faster decay. It has instead a smaller exponent for the range we have constructed,
although it should decay at least as fast as $M^{-2}$, since the quadratic order operator contains the linear order operator. Starting from cubic power, we see a signature of slow operators, where
$\lambda(M)$ decreases faster than $M^{-2}$. How these nonlinear order of energy density operators contribute to slow relaxation remains unclear but our results suggest that they have some nontrivial
consequences.

\subsubsection{Filtered operator}

\begin{figure}
\includegraphics[width=1.0\linewidth]{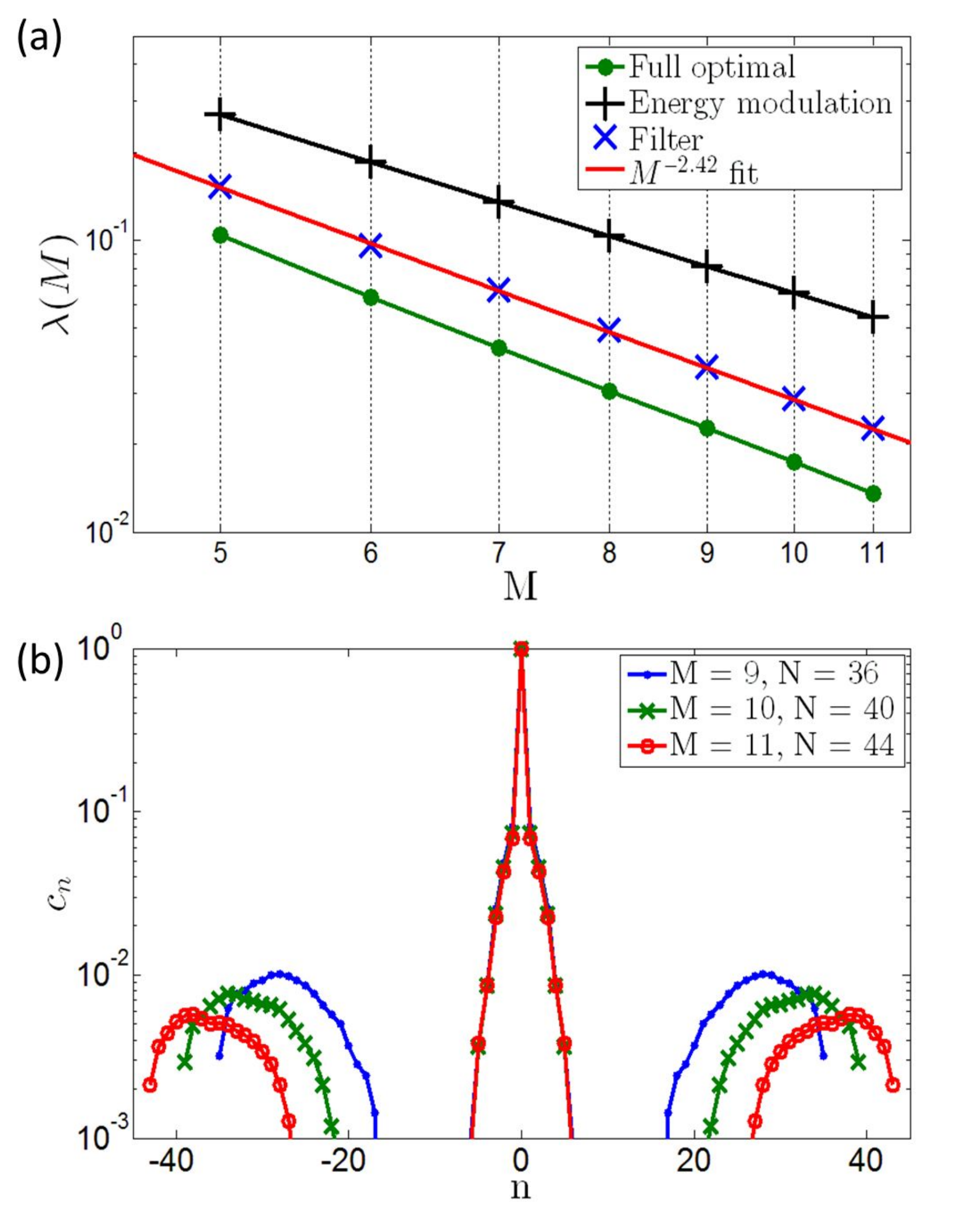}
\centering \caption{(color online) Filtering operator of the Hamiltonian system. (a) $\lambda(M)$ is the minimum value of the square
of the Frobenius norm of the commutator of the filtered operator (Eq. ~\eqref{eq:filter_H}) with Hamiltonian. It decreases faster than the diffusion scaling, $M^{-2}$. Therefore, they are slow operators.
(b) The coefficients of slowest filtered operator. It is highly peaked near $n = 0$ as expected. In addition, they get small but nonnegligible contributions from the later step operators.}
\label{fig:filter_H}
\end{figure}

Another way of constructing slow operators is the filtering of energy modulation. A conventional filtering method (see Eq.\eqref{eq:filter}) is discussed in the main text where we construct the slowest
operator of the Floquet system. In the Hamiltonian case, however, the method should be modified since in principle for any $t\neq 0$, $\exp(-i H t)$ is a global operator that makes a local operator of
range $M$ act on every site. In addition, it turns out that it is not easy to build slow operators for a Hamiltonian system if we start from a single site operator. Instead, we use the cosine modulation of
the energy density operator as starting operator and trace out the external sites to restrict the range of operators to be $M$ at every step. We construct an operator of the following form:
\begin{align}
{\hat{A}_{M,N}} &= \sum_{n=-N}^{N} c_n \hat{a}_{M,n} \label{eq:filter_H}\\
\hat{a}_{M,n} &= \mathrm{tr_{\setminus M}}( e^{i H n\delta t} \hat{a}_{M,n-1} e^{-i H n \delta t}) ~,
\end{align}
where
$\hat{a}_{M,0} $ is the range $M$ cosine modulation of energy density operator, $\mathrm{tr_{\setminus M}}$ traces out the region outside $M$ consecutive spins and $\delta t$ is the small time step. First
we obtain the $2N+1$ $\hat{a}_{M,n}$'s and then we compute the coefficients $c_n$ that minimize the Frobenius norm of the commutator with the Hamiltonian. For a sufficiently small $\delta t$, we can
approximate $e^{i H t}$ with $H$ acting on $M+2$ sites (another site at each edge) with negligible error at each step. We took $\delta t = 0.667$ and checked that results do not change by increasing the
approximate range of $e^{i H t}$. The underlying idea of this construction is that we start from the expected slowest mode (energy modulation) and then ``filter out'' the fast component, if exits, at each
small time evolution. We make the total evolution step number $N$ be proportional to the range $M$.

Figure \ref{fig:filter_H} (a) is the plot of $\lambda(M)$, which is the square of the Frobenius norm of the commutator between the slowest operator found by the filtering scheme and the Hamiltonian. It
decreases faster than energy diffusion and thus we can say that these are indeed slow operators. The coefficients $c_n$ are highly peaked around $n = 0$, which is consistent with the previous observation
that the largest contribution to the slowest mode  comes from the energy density modulation. Nevertheless, we are unable to get the exact slowest operator by varying $\delta n$, $N$, and $\hat{a}_{M,0}$.

\section{Results of another set of parameters and different model Hamiltonians}
\subsection{Another set of parameters}
\begin{figure}
\includegraphics[width=1.0\linewidth]{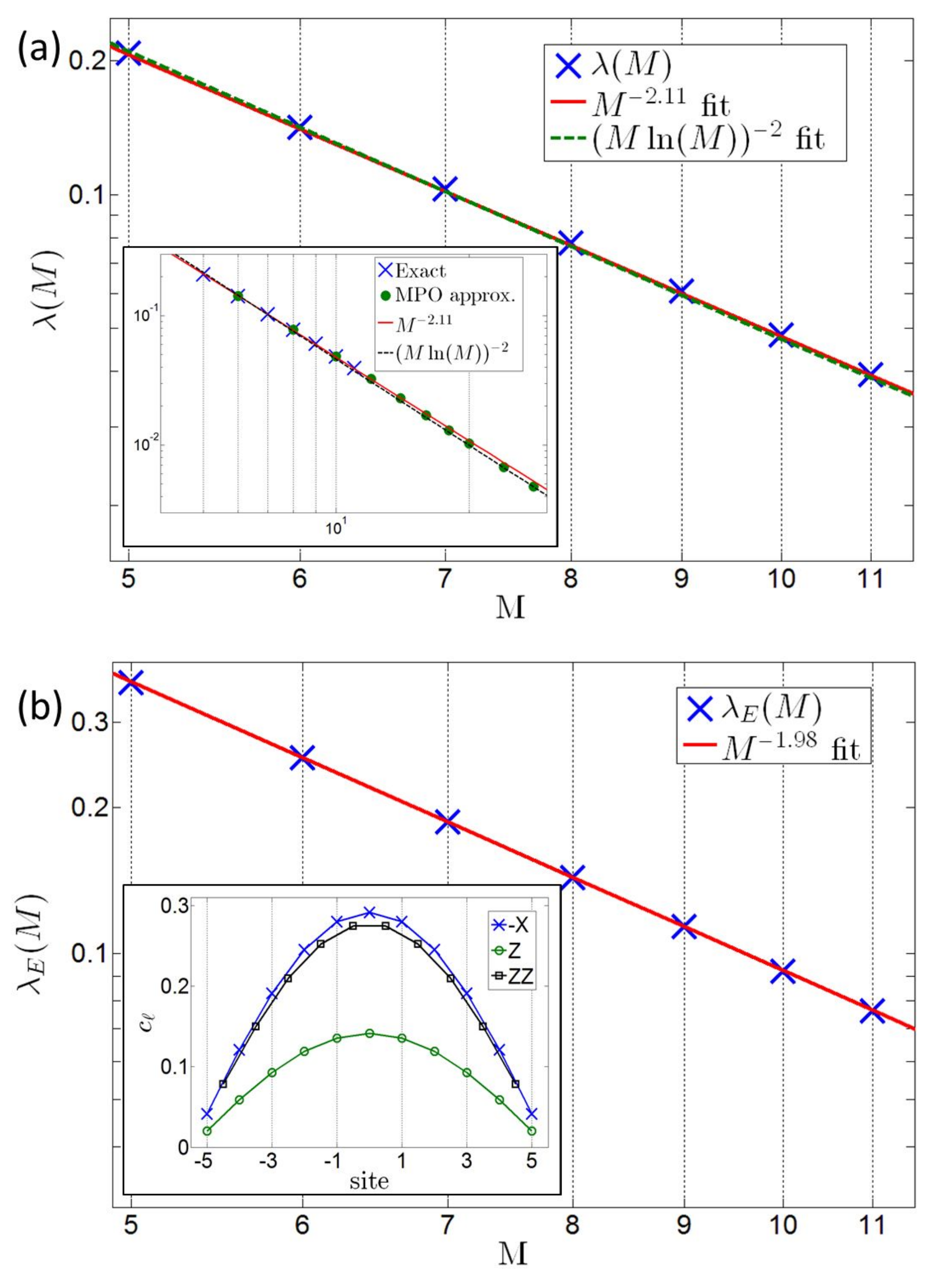}
\centering \caption{(color online) (a) $\lambda(M)$ is the minimum value of the Frobenius norm of the commutator with Hamiltonian.
Here we choose the parameters to be $(g,h) = (-1.05, 0.5)$. $\lambda$ still decreases faster than $1/M^2$. Inset: $\lambda(M)$ vs. $M$ using tensor network method (MPO approximation). The results are
indistinguishable from available exact results. (b) $\lambda_E(M)$ is the same as when we restrict the search space within the terms in the Hamiltonian. As expected, it has $M^{-2}$ scaling and the
structure is cosine modulation (inset).}
\label{fig:hamiltonian_other}
\end{figure}
In this section, we show that the results in the main text do not depend on the parameter choice. We choose another set
of parameters $(g,h) = (-1.05, 0.5)$, which is the parameter choice of Ref.~\onlinecite{Banuls:2011}.

Figure \ref{fig:hamiltonian_other} (a) is the minimum value of Eq.\eqref{eq:minimize} in the main text with the other parameter choice. We can see that it decays faster than $1/M^2$. As is the case of the
parameter choice in the main text, the data can be well-fitted by two methods; a power-law and a logarithmic correction to $1/M^2$. Since the power-law exponent could depend on the parameter choice, we do
not attempt to draw a strong conclusion from this data except that $\lambda(M)$ decreases faster than $1/M^2$, the scaling of the diffusive energy mode.

Figure \ref{fig:hamiltonian_other} (b) is the plot of the minimum value of Eq.\eqref{eq:minimize} in the main text when only terms in Hamiltonian are allowed for two sets of parameters. Unlike the case
where all operators are used, the decay scaling remains the same as $1/M^2$ as expected from the hydrodynamics. Therefore, we again explicitly demonstrate that the longest wavelength energy modulation is
the slowest mode of a conserved quantity.

\subsection{Different model Hamiltonians}
In this section, we consider another nonintegrable Hamiltonian and show that there exist slow operators with nontrivial scaling of $\lambda(M)$.

We consider the $XXZ$ model with fields along $x$ and $z$ directions.
\begin{align}
H = \sum_i \sigma^x_i \sigma^x_{i+1} + \sigma^y_i \sigma^y_{i+1} + J_z \sigma^z_i \sigma^z_{i+1} + g \sigma^x_i +h\sigma^z_i ~.
\label{eq:XXZ}
\end{align}
This model is nonintegrable when at least two of three parameters ($J_z, g, h$) are nonzero. We choose two cases:
$(J_z, g, h) = (0.5, 0.8, 0.4)$ and $(J_z, g, h) = (0, 0.8, 0.4)$.
These choices have fewer discrete symmetries than the other two possibilities; when only $g = 0$, this model conserves total spin along $z$ direction and when only $h = 0$, this model is
symmetric under all spin flip (Ising symmetry).

\begin{figure}
\includegraphics[width=1.0\linewidth]{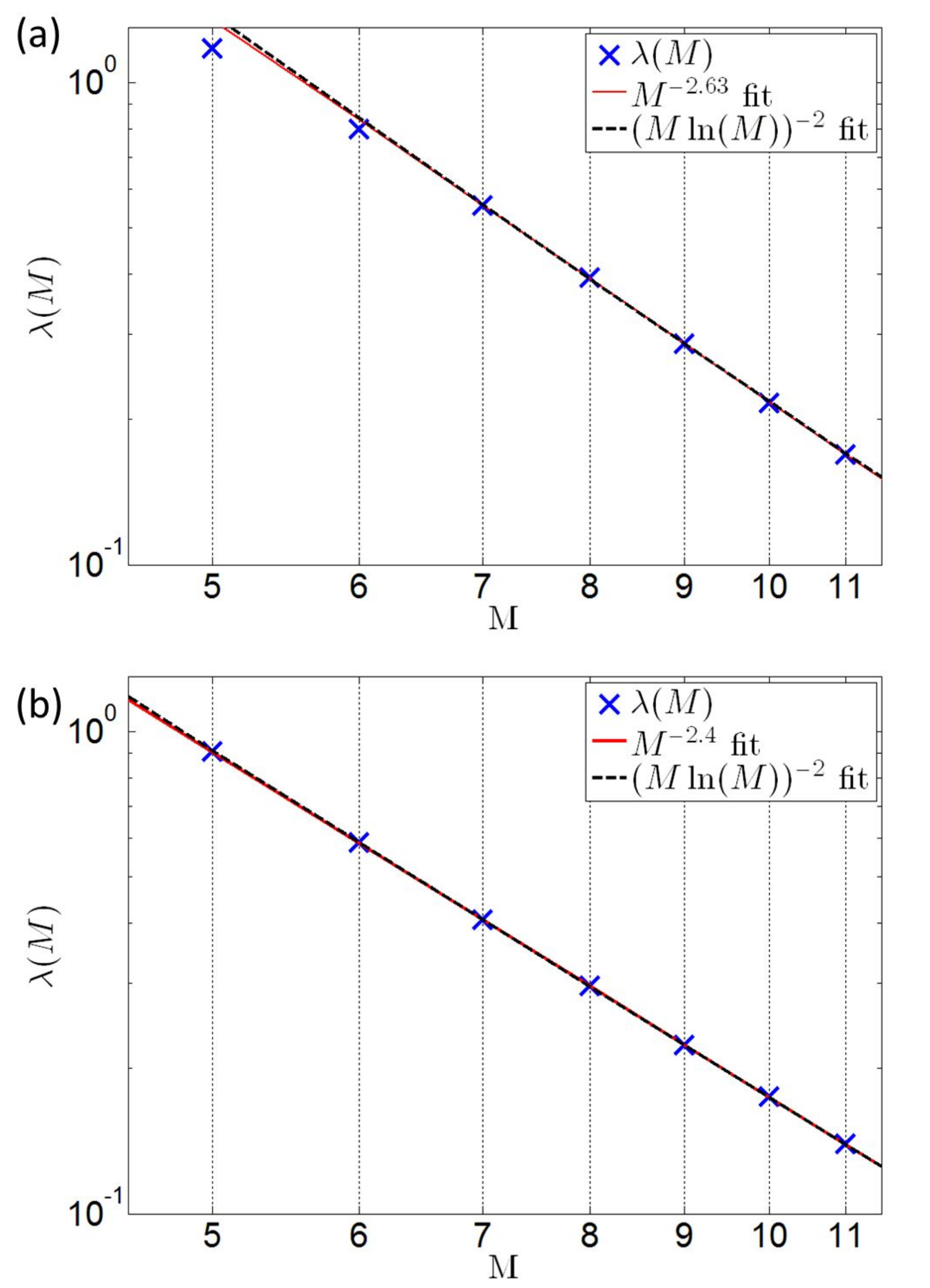}
\centering \caption{(color online) $\lambda(M)$ is the minimum value of the Frobenius norm of the commutator with Hamiltonian, Eq.
~\eqref{eq:XXZ}. (a) $(J_z, g, h) = (0.5, 0.8, 0.4)$. (b) $(J_z, g, h) = (0, 0.8, 0.4)$. Both cases, $\lambda(M)$ decreases faster than $M^{-2}$ and thus there exist nontrivial slow operators. }
\label{fig:other_Ham}
\end{figure}

Figure \ref{fig:other_Ham} is the plot of $\lambda(M)$ vs. $M$, where $\lambda(M)$ is the smallest value of the Frobenius norm of the commutator with the Hamiltonian for an operator of range $M$. We can
clearly see that in both cases $\lambda(M)$ decreases faster than diffusive scaling, $M^{-2}$. Therefore, there exists some slow operator that relaxes slower than diffusion in this nonintegrable model. We
have obtained similar results for the other two possibilities of parameter choices and there we found larger exponent, which may result from existence of other symmetries.

\section{Results of Matrix Product Operators with various bond dimensions}

\begin{figure}
\includegraphics[width=1.0\linewidth]{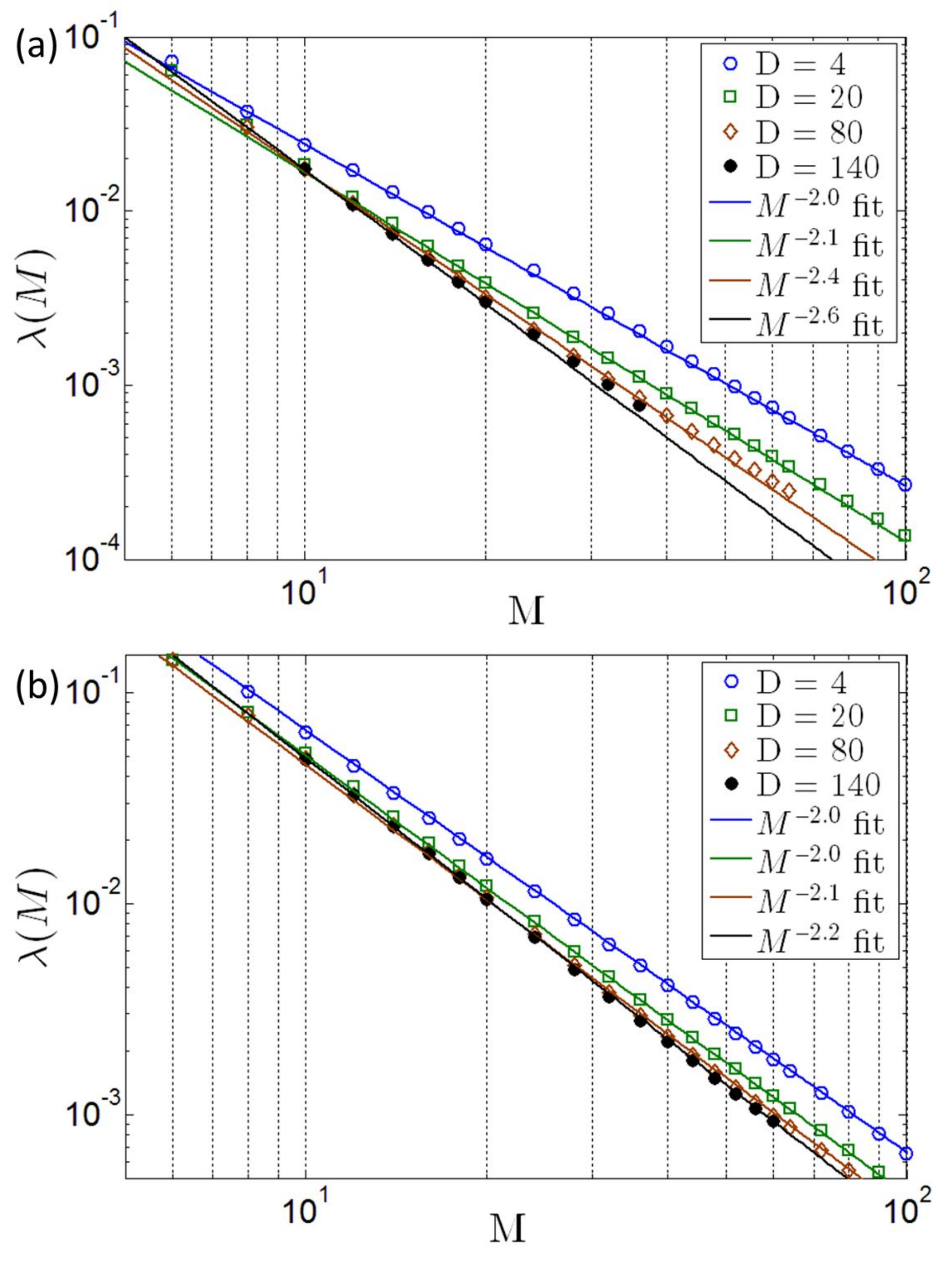}
\centering \caption{(color online) (a) $\lambda(M)$ computed by the matrix product ansatz at fixed bond dimension $D$. Straight
lines are power-law fittings to the $\lambda(M)$'s computed. At $D = 80$, we already see significant deviation from the diffusion scaling. This is a strong evidence that we have operators relaxing slower
than the hydrodynamic mode. (b) Same calculation with a different parameter set; $(g,h) = (-1.05, 0.5)$. Qualitative features are the same. }
\label{fig:MPO_results}
\end{figure}

In this section, we present more results of the matrix product ansatz with various bond dimensions for the Hamiltonian system. For a fixed bond dimension $D$, we compute the matrix product operator (MPO)
that minimizes $\lambda(M)$. This gives another upper bound to the true minimum value of given support $M$. Fig. \ref{fig:MPO_results} shows the results of $D = 4, 20, 80,$ and 140. Since the energy
modulation can be expressed by a matrix product operator with bond dimension 3, operators of bond dimension 4 contain the energy modulation.
This explains what we see in the Fig. \ref{fig:MPO_results}
where $\lambda(M) \sim M^{-2}$ (same behavior as the energy diffusion) at $D = 4$.
For growing bond dimensions, we see increasingly larger deviations from the diffusion scaling ($M^{-2}$).
This supports
our conclusion in the main text that there exists nontrivial slowly relaxing operators in the nonintegralbe Hamiltonian system.

\section{Translation invariant operators}
In the main text, we have studied non-translationally invariant operators in order to compare with the visualized sinusoidal energy modulation.
However, the general
procedure of minimizing the square of the Frobenius norm of the commutator with Hamiltonian is not restricted to non-translationally invariant system.
In addition, translation invariant system is more directly
connected to Ref.~\onlinecite{Banuls:2011}.

First, we define the length $M$ translation invariant operator $\hat{B}_M$ as following:
\begin{align}
\hat{B}_M = \sum_i \hat{A}_{M,i}~,
\end{align}
where $\hat{A}_{M,i}$ is the length $M$ traceless
Hermitian operator supported on sites from $i$ to $i+M-1$. Now, we just search for an optimal operator
$\hat{B}_M$ that minimizes the Frobenius norm of the commutator with Hamiltonian as
Eq.~\eqref{eq:minimize}. What changes here is that once we expand $\hat{A}_{M,i}$ in terms of basis operators as Eq.\eqref{eq:basis_expand},
the denominator in Eq.~\eqref{eq:minimize} becomes different
since basis operators from different sites may no longer be orthogonal.
Therefore, finding the minimum becomes a generalized eigenvalue problem.

\begin{figure}
\includegraphics[width=1.0\linewidth]{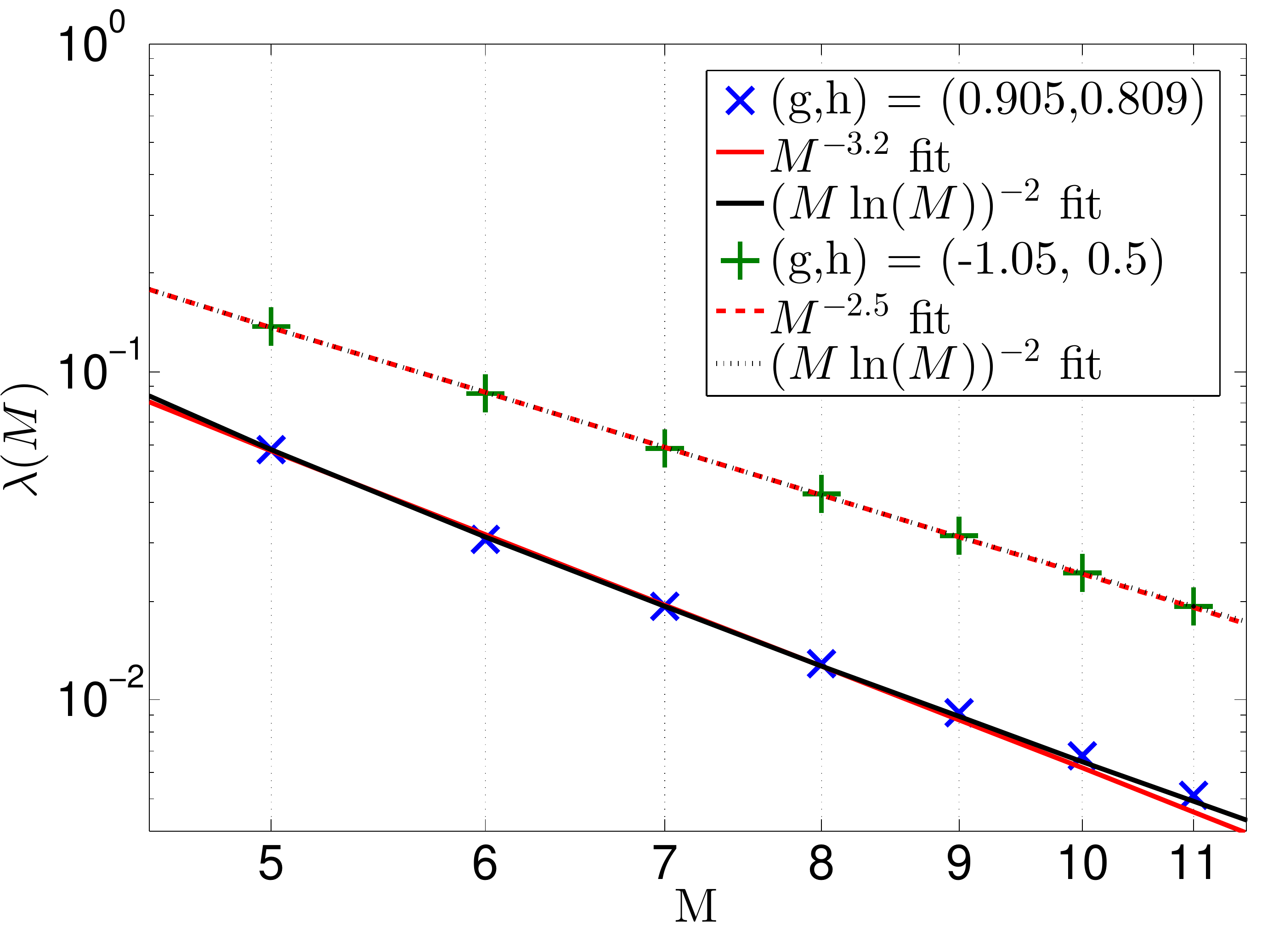}
\centering \caption{(color online) ($\lambda(M)$ of translation invariant operators for two sets of parameter choices. Both
cases, $\lambda(M)$ decreases faster than $M^{-2}$. }
\label{fig:ham_TI}
\end{figure}

Figure \ref{fig:ham_TI} is the results of minimum $\lambda(M)$ for the translation invariant operators. Again, we see that the decay rate of $\lambda(M)$ is faster than $M^{-2}$ although it is not simply
connected to $M^{-2}$ decay of energy modulation.

\section{Existence of Slow Operators for Arbitrary Quantum Circuits}

We now show that a slowly relaxing operator must always exist, on an interval of length $M$ with the relaxation rate going to zero as $M$ gets large.  Let $U$ be the unitary of the quantum circuit
restricted to an interval of length slightly larger than $M$ (in this way, we can consider only finite dimensional spaces).

Let ${\cal E}(\hat{O})$ be a super-operator defined by ${\cal E}(\hat{O})=U\hat{O}U^\dagger$.  The space of operators can be regarded as a vector space, with inner product
$(A,B)={\rm tr}(A^\dagger B)$,
and with ${\cal E}(\hat{O})$ being a linear operator on this space.
The operator ${\cal E}$ is non-Hermitian but it is a normal operator, since its Hermitian conjugate is equal to ${\cal
E}^\dagger(\hat{O})=U^\dagger \hat{O} U$ and ${\cal E}({\cal E}^\dagger(\hat{O}))={\cal E}^\dagger({\cal E}(\hat{O}))=\hat{O}$ and so $[{\cal E},{\cal E}^\dagger]=0$. Let
${\cal E}_h=\frac{1}{2}({\cal E}+{\cal E}^\dagger)$ and ${\cal E}_a=\frac{1}{2i}({\cal E}-{\cal E}^\dagger)$.

We begin by constructing an operator $\hat{O}$ supported on an interval of length $M$ such that ${\cal E}_h(\hat{O})= x \hat{O}+\hat{\epsilon}$ for some $x$ and for $\hat{\epsilon}={\cal O}(1/M)$.  Let $A$
be any traceless operator supported on a single site in the center of the interval.
We consider the Krylov space generated by the vectors $A, {\cal E}_h(A), {\cal E}_h^2(A),...$.  Let $K$ be the number of
vectors we take; $K$ will be proportional to $M$ and will be chosen such that all these operators are supported on the interval of length $M$. Since ${\cal E}_h$ is Hermitian, using the Lanczos procedure
we can write it as a tridiagonal matrix $T$ in this Krylov space.
We claim that there exists a vector $v$ supported on the first $K-1$ vectors $w_1,...,w_{K-1}$ such that $|Tv-xv|^2/|v|^2\leq {\cal
O}(1/K^2)$.
To verify this claim, let $\psi$ be any eigenvector of $T$ with at least half its weight on the first $K/2$ vectors with $\psi=\sum_k a_k v_k$; let $x$ be the corresponding eigenvalue; let
$v=\sum_k a_k (1-k/K) v_k$. The basis in which the matrix is tridiagonal has basis vectors $w_1,w_2,...$, with $w_k$ being in the span of the first $k$ vectors $A, {\cal E}_h(A),...$ Hence, the vector $v$
gives us the desired operator $\hat{O}$.

Now consider the operators
\be
\ hat{O}_{\pm}=(\sqrt{1-x^2} \pm {\cal E}_a) \hat{O},
\ee
with $\hat{O}$ normalized such that $||\hat{O}||=1$.
Note that ${\cal E}(\hat{O}_{\pm})={\cal E}_h(\hat{O}_{\pm})+i {\cal E}_a(\hat{O}_{\pm})=x\hat{O}_{\pm}+i(\sqrt{1-x^2}{\cal E}_a \pm {\cal E}_a^2) \hat{O} + {\cal O}(1/M)$.
Using the fact that ${\cal E}_h^2+{\cal E}_a^2$ is equal to the identity super-operator,
$(\sqrt{1-x^2}{\cal E}_a \pm {\cal E}_a^2) \hat{O}=\sqrt{1-x^2}{\cal E}_a \pm (1-x^2) ) \hat{O} +{\cal O}(1/M)=\pm\sqrt{1-x^2}\hat{O}_{\pm}+{\cal O}(1/M)$. Hence,
\be {\cal E}(\hat{O}_{\pm})=z \hat{O}_{\pm}+{\cal O}(1/M)
\ee
with
\be z=x\pm i \sqrt{1-x^2},
\ee
so that $|z|=1$. At least one of the two operators $\hat{O}_{\pm}$ must have non-negligible norm.  Let $X$ be the corresponding operator,
normalized to have norm $1$.  Hence, we have constructed an operator $X$ supported on the interval of length $M$ such that ${\cal E}(X)=z X + {\cal O}(1/M)$.

This already implies that there is some operator $X$ which is slowly relaxing but perhaps oscillating; i.e., since $X$ is an approximate eigenoperator of ${\cal E}$, if $z=1$ then $X$ changes slowly over
time, while if $z \neq 1$, then the expectation value of $X$ oscillates.

In fact, we can always construct an operator $Y$ which is an approximate eigenoperator of ${\cal E}$.  Here is one way.  Consider many disjoint intervals of length $M$.  Let $X_1,X_2,...$ be the operators
on these intervals with eigenvalues $z_1,z_2,...$  Choose some subset $S$ of these intervals such that the product of the $z_i$ on that subset is close to $1$: $\prod_{i \in S} z_i \approx 1$.  Then, let
$Y=\prod_{i \in S} X_i$.  This requires some analytic estimates to determine the support of $Y$ required: since $Y$ is a product of many operators, the error (in that each $X_i$ is only an approximate
eigenoperator) may add, so the support of $Y$ may scale as a fairly large polynomial in the error.  We leave this estimate for later.

\section{$1/M^2$ scaling of filtered operators}
We show $1/M^2$ scaling of filtered operators of the form Eq. \eqref{eq:filter} if $\mathrm{tr}(\hat{O}U^n \hat{O}U^{-n}) = 0$ for all $n\neq 0$, where
$\hat{O}$ is a traceless Hermitian acting on a single site. First, observe that
\begin{align}
& \frac{\mathrm{tr}([\tilde{A}_N,U][\tilde{A}_N,U])}{\mathrm{tr}(\tilde{A}_N\tilde{A}_N^\dag)} \nonumber\\
&=2 - \bigg(\sum_{n,m = -N}^{N}c_n c_m \mathrm{tr}(U^{n-m-1}\hat{O}U^{-n+m+1}\hat{O} \nonumber\\
&\quad\quad\quad\quad\quad + U^{n-m+1}\hat{O}U^{-n+m-1}\hat{O})\bigg)/\mathrm{tr}(\tilde{A}_N\tilde{A}_N^\dag)~.
\end{align}
Therefore, if $\mathrm{tr}(\hat{O}U^n \hat{O}U^{-n}) = \delta_{n,0}\mathrm{tr}(\hat{O}\hat{O}^\dag)$, the above expression simplifies to
\begin{align}
\frac{\mathrm{tr}([\tilde{A}_N,U][\tilde{A}_N,U])}{\mathrm{tr}(\tilde{A}_N\tilde{A}_N^\dag)}=2\left(1 - \frac{\sum_{n=-N}^{N-1}c_n c_{n+1}}{\sum_{n=-N}^N c_n^2}\right) ~.
\end{align}
This is a trivial
quadratic optimization problem and the solution is $c_n = \cos(n\pi/(2N+2))$.
Therefore, the minimum value $\tilde{\lambda}$ for sufficiently large $N$ is
\begin{align}
\tilde{\lambda} &= \mathrm{min}\left[\frac{\mathrm{tr}([\tilde{A}_N,U][\tilde{A}_N,U])}{\mathrm{tr}(\tilde{A}_N\tilde{A}_N^\dag)} \right] \nonumber\\
&= 2 - 2\cos\left(\frac{\pi}{2N+2}\right) \simeq \frac{\pi^2}{4(N+1)^2} \sim \frac{1}{M^2} ~,
\end{align}
where we used the fact that the support $M = 4N +1 $.

Ref. ~\onlinecite{Kim_ETH} has shown that the $U_F$ (the Floquet operator) thermalizes a local operator at infinite temperature,
and thus $\mathrm{tr}(\hat{O}U^n\hat{O}U^{-n}) = 0$ for a sufficiently large
$n$. For the Floquet system, therefore, the above condition is approximately satisfied for a large enough $N$. Figure \ref{fig:floquet} in the main text shows that for $N$ = 60 and 100, the form is very
close to the cosine and the scaling for large $M = 2N +1$ closely follows $M^{-2}$ scaling. For a general random circuit, Figure \ref{fig:random_circuit} again shows $M^{-2}$ scaling. The structure of
$\{c_n\}$ is found to follow cosine modulation similar to the Floquet case.

One example of quantum circuits that satisfies the condition that $\mathrm{tr}(\hat{O}U^{n} \hat{O}U^{-n}) = 0$ for all $n\neq0$ is the swap operator $U_{sw}$.
\begin{align}
U_{sw} = \prod_n V_{2n,2n+1}
\prod_m V_{2m-1,2m} ~,
\end{align}
where $V_{x,y}$ swaps the spins ${\bf S}_x$ and ${\bf S}_y$; $V_{x,y}|{\bf S}_x,{\bf S}_y\rangle = |{\bf S}_y, {\bf S}_x\rangle$. Then, for an $\hat{O}$ acting on site 0,
$U_{sw}$ moves $\hat{O}$ to the site $-2n$ and $U_{sw}^{-n}$ moves $\hat{O}$ to the site $2n$ and thus the condition is satisfied. In this case, we can write every step analytically and prove $1/M^2$
scaling and the cosine modulation. In a generic quantum circuit, however, these features are seen only for sufficiently large $M$. Furthermore, in a translation invariant system, $U_{sw}$ has an exact
local conservation law. For instance, any translation invariant single site operator commutes with $U_{sw}$ and is thus conserved.

In the main text, we have connected $\lambda(M)$ to the thermalization time scale by $1/\tau \sim \sqrt{\lambda(M)}$ (Eq.\eqref{eq:floquet_timescale}). At first glance, $1/\tau \sim \sqrt{\lambda(M)} \sim
1/M$ may seem trivial for an operator of support $M$ by Lieb-Robinson bound type argument, but an important point here is that we computed the relaxation of an operator by the Frobenius norm, not the
operator norm by which the Lieb-Robinson bound has been computed \cite{Lieb:1972}.

\bibliography{slow_op_bibl}

\begin{thebibliography}{34}%
\makeatletter
\providecommand \@ifxundefined [1]{%
 \@ifx{#1\undefined}
}%
\providecommand \@ifnum [1]{%
 \ifnum #1\expandafter \@firstoftwo
 \else \expandafter \@secondoftwo
 \fi
}%
\providecommand \@ifx [1]{%
 \ifx #1\expandafter \@firstoftwo
 \else \expandafter \@secondoftwo
 \fi
}%
\providecommand \natexlab [1]{#1}%
\providecommand \enquote  [1]{``#1''}%
\providecommand \bibnamefont  [1]{#1}%
\providecommand \bibfnamefont [1]{#1}%
\providecommand \citenamefont [1]{#1}%
\providecommand \href@noop [0]{\@secondoftwo}%
\providecommand \href [0]{\begingroup \@sanitize@url \@href}%
\providecommand \@href[1]{\@@startlink{#1}\@@href}%
\providecommand \@@href[1]{\endgroup#1\@@endlink}%
\providecommand \@sanitize@url [0]{\catcode `\\12\catcode `\$12\catcode
  `\&12\catcode `\#12\catcode `\^12\catcode `\_12\catcode `\%12\relax}%
\providecommand \@@startlink[1]{}%
\providecommand \@@endlink[0]{}%
\providecommand \url  [0]{\begingroup\@sanitize@url \@url }%
\providecommand \@url [1]{\endgroup\@href {#1}{\urlprefix }}%
\providecommand \urlprefix  [0]{URL }%
\providecommand \Eprint [0]{\href }%
\providecommand \doibase [0]{http://dx.doi.org/}%
\providecommand \selectlanguage [0]{\@gobble}%
\providecommand \bibinfo  [0]{\@secondoftwo}%
\providecommand \bibfield  [0]{\@secondoftwo}%
\providecommand \translation [1]{[#1]}%
\providecommand \BibitemOpen [0]{}%
\providecommand \bibitemStop [0]{}%
\providecommand \bibitemNoStop [0]{.\EOS\space}%
\providecommand \EOS [0]{\spacefactor3000\relax}%
\providecommand \BibitemShut  [1]{\csname bibitem#1\endcsname}%
\let\auto@bib@innerbib\@empty
\bibitem [{\citenamefont {Deutsch}(1991)}]{Deutsch:1991}%
  \BibitemOpen
  \bibfield  {author} {\bibinfo {author} {\bibfnamefont {J.~M.}\ \bibnamefont
  {Deutsch}},\ }\bibfield  {title} {\enquote {\bibinfo {title} {Quantum
  statistical mechanics in a closed system},}\ }\href {\doibase
  10.1103/PhysRevA.43.2046} {\bibfield  {journal} {\bibinfo  {journal} {Phys.
  Rev. A}\ }\textbf {\bibinfo {volume} {43}},\ \bibinfo {pages} {2046--2049}
  (\bibinfo {year} {1991})}\BibitemShut {NoStop}%
\bibitem [{\citenamefont {Srednicki}(1994)}]{Srednicki:1994}%
  \BibitemOpen
  \bibfield  {author} {\bibinfo {author} {\bibfnamefont {M}~\bibnamefont
  {Srednicki}},\ }\bibfield  {title} {\enquote {\bibinfo {title} {Chaos and
  quantum thermalization},}\ }\href {\doibase 10.1103/PhysRevE.50.888}
  {\bibfield  {journal} {\bibinfo  {journal} {Phys. Rev. E}\ }\textbf {\bibinfo
  {volume} {50}},\ \bibinfo {pages} {888--901} (\bibinfo {year}
  {1994})}\BibitemShut {NoStop}%
\bibitem [{\citenamefont {Rigol}\ \emph {et~al.}(2008)\citenamefont {Rigol},
  \citenamefont {Dunjko},\ and\ \citenamefont {Olshanii}}]{Rigol:2008}%
  \BibitemOpen
  \bibfield  {author} {\bibinfo {author} {\bibfnamefont {Marcos}\ \bibnamefont
  {Rigol}}, \bibinfo {author} {\bibfnamefont {Vanja}\ \bibnamefont {Dunjko}}, \
  and\ \bibinfo {author} {\bibfnamefont {Maxim}\ \bibnamefont {Olshanii}},\
  }\bibfield  {title} {\enquote {\bibinfo {title} {Thermalization and its
  mechanism for generic isolated quantum systems},}\ }\href
  {http://dx.doi.org/10.1038/nature06838} {\bibfield  {journal} {\bibinfo
  {journal} {Nature}\ }\textbf {\bibinfo {volume} {452}},\ \bibinfo {pages}
  {854--858} (\bibinfo {year} {2008})}\BibitemShut {NoStop}%
\bibitem [{\citenamefont {Linden}\ \emph {et~al.}(2009)\citenamefont {Linden},
  \citenamefont {Popescu}, \citenamefont {Short},\ and\ \citenamefont
  {Winter}}]{Linden:2009}%
  \BibitemOpen
  \bibfield  {author} {\bibinfo {author} {\bibfnamefont {Noah}\ \bibnamefont
  {Linden}}, \bibinfo {author} {\bibfnamefont {Sandu}\ \bibnamefont {Popescu}},
  \bibinfo {author} {\bibfnamefont {Anthony~J.}\ \bibnamefont {Short}}, \ and\
  \bibinfo {author} {\bibfnamefont {Andreas}\ \bibnamefont {Winter}},\
  }\bibfield  {title} {\enquote {\bibinfo {title} {Quantum mechanical evolution
  towards thermal equilibrium},}\ }\href {\doibase 10.1103/PhysRevE.79.061103}
  {\bibfield  {journal} {\bibinfo  {journal} {Phys. Rev. E}\ }\textbf {\bibinfo
  {volume} {79}},\ \bibinfo {pages} {061103} (\bibinfo {year}
  {2009})}\BibitemShut {NoStop}%
\bibitem [{\citenamefont {Rigol}\ \emph {et~al.}(2007)\citenamefont {Rigol},
  \citenamefont {Dunjko}, \citenamefont {Yurovsky},\ and\ \citenamefont
  {Olshanii}}]{Rigol:2007}%
  \BibitemOpen
  \bibfield  {author} {\bibinfo {author} {\bibfnamefont {Marcos}\ \bibnamefont
  {Rigol}}, \bibinfo {author} {\bibfnamefont {Vanja}\ \bibnamefont {Dunjko}},
  \bibinfo {author} {\bibfnamefont {Vladimir}\ \bibnamefont {Yurovsky}}, \ and\
  \bibinfo {author} {\bibfnamefont {Maxim}\ \bibnamefont {Olshanii}},\
  }\bibfield  {title} {\enquote {\bibinfo {title} {Relaxation in a completely
  integrable many-body quantum system: An ab initio study of the dynamics of
  the highly excited states of 1d lattice hard-core bosons},}\ }\href {\doibase
  10.1103/PhysRevLett.98.050405} {\bibfield  {journal} {\bibinfo  {journal}
  {Phys. Rev. Lett.}\ }\textbf {\bibinfo {volume} {98}},\ \bibinfo {pages}
  {050405} (\bibinfo {year} {2007})}\BibitemShut {NoStop}%
\bibitem [{\citenamefont {Calabrese}\ \emph {et~al.}(2011)\citenamefont
  {Calabrese}, \citenamefont {Essler},\ and\ \citenamefont
  {Fagotti}}]{Calabrese:2011}%
  \BibitemOpen
  \bibfield  {author} {\bibinfo {author} {\bibfnamefont {Pasquale}\
  \bibnamefont {Calabrese}}, \bibinfo {author} {\bibfnamefont {Fabian H.~L.}\
  \bibnamefont {Essler}}, \ and\ \bibinfo {author} {\bibfnamefont {Maurizio}\
  \bibnamefont {Fagotti}},\ }\bibfield  {title} {\enquote {\bibinfo {title}
  {Quantum quench in the transverse-field ising chain},}\ }\href {\doibase
  10.1103/PhysRevLett.106.227203} {\bibfield  {journal} {\bibinfo  {journal}
  {Phys. Rev. Lett.}\ }\textbf {\bibinfo {volume} {106}},\ \bibinfo {pages}
  {227203} (\bibinfo {year} {2011})}\BibitemShut {NoStop}%
\bibitem [{\citenamefont {Gogolin}\ \emph {et~al.}(2011)\citenamefont
  {Gogolin}, \citenamefont {M\"uller},\ and\ \citenamefont
  {Eisert}}]{Gogolin:2011}%
  \BibitemOpen
  \bibfield  {author} {\bibinfo {author} {\bibfnamefont {Christian}\
  \bibnamefont {Gogolin}}, \bibinfo {author} {\bibfnamefont {Markus~P.}\
  \bibnamefont {M\"uller}}, \ and\ \bibinfo {author} {\bibfnamefont {Jens}\
  \bibnamefont {Eisert}},\ }\bibfield  {title} {\enquote {\bibinfo {title}
  {Absence of thermalization in nonintegrable systems},}\ }\href {\doibase
  10.1103/PhysRevLett.106.040401} {\bibfield  {journal} {\bibinfo  {journal}
  {Phys. Rev. Lett.}\ }\textbf {\bibinfo {volume} {106}},\ \bibinfo {pages}
  {040401} (\bibinfo {year} {2011})}\BibitemShut {NoStop}%
\bibitem [{\citenamefont {Fagotti}\ \emph {et~al.}(2014)\citenamefont
  {Fagotti}, \citenamefont {Collura}, \citenamefont {Essler},\ and\
  \citenamefont {Calabrese}}]{Fagotti:2014}%
  \BibitemOpen
  \bibfield  {author} {\bibinfo {author} {\bibfnamefont {Maurizio}\
  \bibnamefont {Fagotti}}, \bibinfo {author} {\bibfnamefont {Mario}\
  \bibnamefont {Collura}}, \bibinfo {author} {\bibfnamefont {Fabian H.~L.}\
  \bibnamefont {Essler}}, \ and\ \bibinfo {author} {\bibfnamefont {Pasquale}\
  \bibnamefont {Calabrese}},\ }\bibfield  {title} {\enquote {\bibinfo {title}
  {Relaxation after quantum quenches in the spin-$\frac{1}{2}$ heisenberg xxz
  chain},}\ }\href {\doibase 10.1103/PhysRevB.89.125101} {\bibfield  {journal}
  {\bibinfo  {journal} {Phys. Rev. B}\ }\textbf {\bibinfo {volume} {89}},\
  \bibinfo {pages} {125101} (\bibinfo {year} {2014})}\BibitemShut {NoStop}%
\bibitem [{\citenamefont {Wouters}\ \emph {et~al.}(2014)\citenamefont
  {Wouters}, \citenamefont {De~Nardis}, \citenamefont {Brockmann},
  \citenamefont {Fioretto}, \citenamefont {Rigol},\ and\ \citenamefont
  {Caux}}]{Wouters:2014}%
  \BibitemOpen
  \bibfield  {author} {\bibinfo {author} {\bibfnamefont {B.}~\bibnamefont
  {Wouters}}, \bibinfo {author} {\bibfnamefont {J.}~\bibnamefont {De~Nardis}},
  \bibinfo {author} {\bibfnamefont {M.}~\bibnamefont {Brockmann}}, \bibinfo
  {author} {\bibfnamefont {D.}~\bibnamefont {Fioretto}}, \bibinfo {author}
  {\bibfnamefont {M.}~\bibnamefont {Rigol}}, \ and\ \bibinfo {author}
  {\bibfnamefont {J.-S.}\ \bibnamefont {Caux}},\ }\bibfield  {title} {\enquote
  {\bibinfo {title} {Quenching the anisotropic heisenberg chain: Exact solution
  and generalized gibbs ensemble predictions},}\ }\href {\doibase
  10.1103/PhysRevLett.113.117202} {\bibfield  {journal} {\bibinfo  {journal}
  {Phys. Rev. Lett.}\ }\textbf {\bibinfo {volume} {113}},\ \bibinfo {pages}
  {117202} (\bibinfo {year} {2014})}\BibitemShut {NoStop}%
\bibitem [{\citenamefont {Pozsgay}\ \emph {et~al.}(2014)\citenamefont
  {Pozsgay}, \citenamefont {Mesty\'an}, \citenamefont {Werner}, \citenamefont
  {Kormos}, \citenamefont {Zar\'and},\ and\ \citenamefont
  {Tak\'acs}}]{Pozsgay:2014}%
  \BibitemOpen
  \bibfield  {author} {\bibinfo {author} {\bibfnamefont {B.}~\bibnamefont
  {Pozsgay}}, \bibinfo {author} {\bibfnamefont {M.}~\bibnamefont {Mesty\'an}},
  \bibinfo {author} {\bibfnamefont {M.~A.}\ \bibnamefont {Werner}}, \bibinfo
  {author} {\bibfnamefont {M.}~\bibnamefont {Kormos}}, \bibinfo {author}
  {\bibfnamefont {G.}~\bibnamefont {Zar\'and}}, \ and\ \bibinfo {author}
  {\bibfnamefont {G.}~\bibnamefont {Tak\'acs}},\ }\bibfield  {title} {\enquote
  {\bibinfo {title} {Correlations after quantum quenches in the $xxz$ spin
  chain: Failure of the generalized gibbs ensemble},}\ }\href {\doibase
  10.1103/PhysRevLett.113.117203} {\bibfield  {journal} {\bibinfo  {journal}
  {Phys. Rev. Lett.}\ }\textbf {\bibinfo {volume} {113}},\ \bibinfo {pages}
  {117203} (\bibinfo {year} {2014})}\BibitemShut {NoStop}%
\bibitem [{\citenamefont {Goldstein}\ and\ \citenamefont
  {Andrei}(2014{\natexlab{a}})}]{Goldstein:2014-2}%
  \BibitemOpen
  \bibfield  {author} {\bibinfo {author} {\bibfnamefont {G.}~\bibnamefont
  {Goldstein}}\ and\ \bibinfo {author} {\bibfnamefont {N}~\bibnamefont
  {Andrei}},\ }\bibfield  {title} {\enquote {\bibinfo {title} {{Failure of the
  GGE hypothesis for integrable models with bound states}},}\ }\href@noop {}
  {\bibfield  {journal} {\bibinfo  {journal} {ArXiv e-prints}\ } (\bibinfo
  {year} {2014}{\natexlab{a}})},\ \Eprint {http://arxiv.org/abs/1405.4224v2}
  {arXiv:1405.4224v2 [cond-mat.quantumgas]} \BibitemShut {NoStop}%
\bibitem [{\citenamefont {Mestyan}\ \emph {et~al.}(2014)\citenamefont
  {Mestyan}, \citenamefont {Pozsgay}, \citenamefont {Takacs},\ and\
  \citenamefont {Werner}}]{Mestyan:2014}%
  \BibitemOpen
  \bibfield  {author} {\bibinfo {author} {\bibfnamefont {M.}~\bibnamefont
  {Mestyan}}, \bibinfo {author} {\bibfnamefont {B.}~\bibnamefont {Pozsgay}},
  \bibinfo {author} {\bibfnamefont {G.}~\bibnamefont {Takacs}}, \ and\ \bibinfo
  {author} {\bibfnamefont {M.A.}\ \bibnamefont {Werner}},\ }\bibfield  {title}
  {\enquote {\bibinfo {title} {{Quenching the XXZ spin chain: quench action
  approach versus generalized Gibbs ensemble}},}\ }\href@noop {} {\bibfield
  {journal} {\bibinfo  {journal} {ArXiv e-prints}\ } (\bibinfo {year}
  {2014})},\ \Eprint {http://arxiv.org/abs/1412.4787v1} {arXiv:1412.4787v1
  [cond-mat.stat-mech]} \BibitemShut {NoStop}%
\bibitem [{\citenamefont {Polkovnikov}\ \emph {et~al.}(2011)\citenamefont
  {Polkovnikov}, \citenamefont {Sengupta}, \citenamefont {Silva},\ and\
  \citenamefont {Vengalattore}}]{Polkovnikov:2011}%
  \BibitemOpen
  \bibfield  {author} {\bibinfo {author} {\bibfnamefont {Anatoli}\ \bibnamefont
  {Polkovnikov}}, \bibinfo {author} {\bibfnamefont {Krishnendu}\ \bibnamefont
  {Sengupta}}, \bibinfo {author} {\bibfnamefont {Alessandro}\ \bibnamefont
  {Silva}}, \ and\ \bibinfo {author} {\bibfnamefont {Mukund}\ \bibnamefont
  {Vengalattore}},\ }\bibfield  {title} {\enquote {\bibinfo {title}
  {Colloquium},}\ }\href {\doibase 10.1103/RevModPhys.83.863} {\bibfield
  {journal} {\bibinfo  {journal} {Rev. Mod. Phys.}\ }\textbf {\bibinfo {volume}
  {83}},\ \bibinfo {pages} {863--883} (\bibinfo {year} {2011})}\BibitemShut
  {NoStop}%
\bibitem [{\citenamefont {Yukalov}(2011)}]{Yukalov:2011}%
  \BibitemOpen
  \bibfield  {author} {\bibinfo {author} {\bibfnamefont {V.I.}\ \bibnamefont
  {Yukalov}},\ }\bibfield  {title} {\enquote {\bibinfo {title} {Equilibration
  and thermalization in finite quantum systems},}\ }\href {\doibase
  10.1002/lapl.201110002} {\bibfield  {journal} {\bibinfo  {journal} {Laser
  Phys. Lett.}\ }\textbf {\bibinfo {volume} {8}},\ \bibinfo {pages} {485}
  (\bibinfo {year} {2011})}\BibitemShut {NoStop}%
\bibitem [{\citenamefont {Santos}\ and\ \citenamefont
  {Rigol}(2010)}]{Santos:2010}%
  \BibitemOpen
  \bibfield  {author} {\bibinfo {author} {\bibfnamefont {Lea~F.}\ \bibnamefont
  {Santos}}\ and\ \bibinfo {author} {\bibfnamefont {Marcos}\ \bibnamefont
  {Rigol}},\ }\bibfield  {title} {\enquote {\bibinfo {title} {Localization and
  the effects of symmetries in the thermalization properties of one-dimensional
  quantum systems},}\ }\href {\doibase 10.1103/PhysRevE.82.031130} {\bibfield
  {journal} {\bibinfo  {journal} {Phys. Rev. E}\ }\textbf {\bibinfo {volume}
  {82}},\ \bibinfo {pages} {031130} (\bibinfo {year} {2010})}\BibitemShut
  {NoStop}%
\bibitem [{\citenamefont {Rigol}\ and\ \citenamefont
  {Srednicki}(2012)}]{Rigol:2012}%
  \BibitemOpen
  \bibfield  {author} {\bibinfo {author} {\bibfnamefont {M.}~\bibnamefont
  {Rigol}}\ and\ \bibinfo {author} {\bibfnamefont {M.}~\bibnamefont
  {Srednicki}},\ }\bibfield  {title} {\enquote {\bibinfo {title} {Alternatives
  to eigenstate thermalization},}\ }\href {\doibase
  10.1103/PhysRevLett.108.110601} {\bibfield  {journal} {\bibinfo  {journal}
  {Phys. Rev. Lett.}\ }\textbf {\bibinfo {volume} {108}},\ \bibinfo {pages}
  {110601} (\bibinfo {year} {2012})}\BibitemShut {NoStop}%
\bibitem [{\citenamefont {Kruczenski}\ and\ \citenamefont
  {Khlebnikov}(2013)}]{Kruczenski:2013}%
  \BibitemOpen
  \bibfield  {author} {\bibinfo {author} {\bibfnamefont {M}~\bibnamefont
  {Kruczenski}}\ and\ \bibinfo {author} {\bibfnamefont {S.}~\bibnamefont
  {Khlebnikov}},\ }\bibfield  {title} {\enquote {\bibinfo {title}
  {Thermalization of isolated quantum systems},}\ }\href@noop {} {\bibfield
  {journal} {\bibinfo  {journal} {ArXiv e-prints}\ } (\bibinfo {year}
  {2013})},\ \Eprint {http://arxiv.org/abs/1312.4612v2} {arXiv:1312.4612v2}
  \BibitemShut {NoStop}%
\bibitem [{\citenamefont {Beugeling}\ \emph {et~al.}(2014)\citenamefont
  {Beugeling}, \citenamefont {Moessner},\ and\ \citenamefont
  {Haque}}]{Beugeling:2014}%
  \BibitemOpen
  \bibfield  {author} {\bibinfo {author} {\bibfnamefont {W.}~\bibnamefont
  {Beugeling}}, \bibinfo {author} {\bibfnamefont {R.}~\bibnamefont {Moessner}},
  \ and\ \bibinfo {author} {\bibfnamefont {Masudul}\ \bibnamefont {Haque}},\
  }\bibfield  {title} {\enquote {\bibinfo {title} {Finite-size scaling of
  eigenstate thermalization},}\ }\href {\doibase 10.1103/PhysRevE.89.042112}
  {\bibfield  {journal} {\bibinfo  {journal} {Phys. Rev. E}\ }\textbf {\bibinfo
  {volume} {89}},\ \bibinfo {pages} {042112} (\bibinfo {year}
  {2014})}\BibitemShut {NoStop}%
\bibitem [{\citenamefont {Sorg}\ \emph {et~al.}(2014)\citenamefont {Sorg},
  \citenamefont {Vidmar}, \citenamefont {Pollet},\ and\ \citenamefont
  {Heidrich-Meisner}}]{Sorg:2014}%
  \BibitemOpen
  \bibfield  {author} {\bibinfo {author} {\bibfnamefont {S.}~\bibnamefont
  {Sorg}}, \bibinfo {author} {\bibfnamefont {L.}~\bibnamefont {Vidmar}},
  \bibinfo {author} {\bibfnamefont {L.}~\bibnamefont {Pollet}}, \ and\ \bibinfo
  {author} {\bibfnamefont {F.}~\bibnamefont {Heidrich-Meisner}},\ }\bibfield
  {title} {\enquote {\bibinfo {title} {Relaxation and thermalization in the
  one-dimensional bose-hubbard model: A case study for the interaction quantum
  quench from the atomic limit},}\ }\href {\doibase 10.1103/PhysRevA.90.033606}
  {\bibfield  {journal} {\bibinfo  {journal} {Phys. Rev. A}\ }\textbf {\bibinfo
  {volume} {90}},\ \bibinfo {pages} {033606} (\bibinfo {year}
  {2014})}\BibitemShut {NoStop}%
\bibitem [{\citenamefont {Kim}\ \emph {et~al.}(2014)\citenamefont {Kim},
  \citenamefont {Ikeda},\ and\ \citenamefont {Huse}}]{Kim_ETH}%
  \BibitemOpen
  \bibfield  {author} {\bibinfo {author} {\bibfnamefont {Hyungwon}\
  \bibnamefont {Kim}}, \bibinfo {author} {\bibfnamefont {Tatsuhiko~N.}\
  \bibnamefont {Ikeda}}, \ and\ \bibinfo {author} {\bibfnamefont {David~A.}\
  \bibnamefont {Huse}},\ }\bibfield  {title} {\enquote {\bibinfo {title}
  {Testing whether all eigenstates obey the eigenstate thermalization
  hypothesis},}\ }\href {\doibase 10.1103/PhysRevE.90.052105} {\bibfield
  {journal} {\bibinfo  {journal} {Phys. Rev. E}\ }\textbf {\bibinfo {volume}
  {90}},\ \bibinfo {pages} {052105} (\bibinfo {year} {2014})}\BibitemShut
  {NoStop}%
\bibitem [{\citenamefont {Goldstein}\ and\ \citenamefont
  {Andrei}(2014{\natexlab{b}})}]{Goldstein:2014}%
  \BibitemOpen
  \bibfield  {author} {\bibinfo {author} {\bibfnamefont {G.}~\bibnamefont
  {Goldstein}}\ and\ \bibinfo {author} {\bibfnamefont {N}~\bibnamefont
  {Andrei}},\ }\bibfield  {title} {\enquote {\bibinfo {title} {{Stron
  eigenstate thermalization hypothesis}},}\ }\href@noop {} {\bibfield
  {journal} {\bibinfo  {journal} {ArXiv e-prints}\ } (\bibinfo {year}
  {2014}{\natexlab{b}})},\ \Eprint {http://arxiv.org/abs/1408.3589v1}
  {arXiv:1408.3589v1 [cond-mat.stat-mech]} \BibitemShut {NoStop}%
\bibitem [{\citenamefont {Ba\~nuls}\ \emph {et~al.}(2011)\citenamefont
  {Ba\~nuls}, \citenamefont {Cirac},\ and\ \citenamefont
  {Hastings}}]{Banuls:2011}%
  \BibitemOpen
  \bibfield  {author} {\bibinfo {author} {\bibfnamefont {M.~C.}\ \bibnamefont
  {Ba\~nuls}}, \bibinfo {author} {\bibfnamefont {J.~I.}\ \bibnamefont {Cirac}},
  \ and\ \bibinfo {author} {\bibfnamefont {M.~B.}\ \bibnamefont {Hastings}},\
  }\bibfield  {title} {\enquote {\bibinfo {title} {Strong and weak
  thermalization of infinite nonintegrable quantum systems},}\ }\href {\doibase
  10.1103/PhysRevLett.106.050405} {\bibfield  {journal} {\bibinfo  {journal}
  {Phys. Rev. Lett.}\ }\textbf {\bibinfo {volume} {106}},\ \bibinfo {pages}
  {050405} (\bibinfo {year} {2011})}\BibitemShut {NoStop}%
\bibitem [{\citenamefont {Brandao}\ \emph {et~al.}(2012)\citenamefont
  {Brandao}, \citenamefont {Harrow},\ and\ \citenamefont
  {Horodecki}}]{Brandao:2012}%
  \BibitemOpen
  \bibfield  {author} {\bibinfo {author} {\bibfnamefont {F.~G.~S.~L.}\
  \bibnamefont {Brandao}}, \bibinfo {author} {\bibfnamefont {A.}~\bibnamefont
  {Harrow}}, \ and\ \bibinfo {author} {\bibfnamefont {M.}~\bibnamefont
  {Horodecki}},\ }\bibfield  {title} {\enquote {\bibinfo {title} {{Local random
  quantum circuits are approximate polynomial-designs}},}\ }\href@noop {}
  {\bibfield  {journal} {\bibinfo  {journal} {ArXiv e-prints}\ } (\bibinfo
  {year} {2012})},\ \Eprint {http://arxiv.org/abs/1208.0692v1}
  {arXiv:1208.0692v1 [quant-ph]} \BibitemShut {NoStop}%
\bibitem [{\citenamefont {Kim}\ and\ \citenamefont {Huse}(2013)}]{Kim:2013}%
  \BibitemOpen
  \bibfield  {author} {\bibinfo {author} {\bibfnamefont {Hyungwon}\
  \bibnamefont {Kim}}\ and\ \bibinfo {author} {\bibfnamefont {David~A.}\
  \bibnamefont {Huse}},\ }\bibfield  {title} {\enquote {\bibinfo {title}
  {Ballistic spreading of entanglement in a diffusive nonintegrable system},}\
  }\href {\doibase 10.1103/PhysRevLett.111.127205} {\bibfield  {journal}
  {\bibinfo  {journal} {Phys. Rev. Lett.}\ }\textbf {\bibinfo {volume} {111}},\
  \bibinfo {pages} {127205} (\bibinfo {year} {2013})}\BibitemShut {NoStop}%
\bibitem [{Note1()}]{Note1}%
  \BibitemOpen
  \bibinfo {note} {We have also performed similar analysis in the infinite
  chain for translationally invariant operators where the action of the local
  operator is the same on all sites. Most of our main results (Figures \ref
  {fig:hamiltonian} (a), \ref {fig:floquet} (a)) remain true. Since it is
  easier to visualize local operators and to directly compare with diffusive
  energy mode for translationally non-invariant operators, we present the
  results on an infinite chain where the local operator is placed on $M$
  specific consecutive sites. See the Appendix.}\BibitemShut {Stop}%
\bibitem [{\citenamefont {Pirvu}\ \emph {et~al.}(2010)\citenamefont {Pirvu},
  \citenamefont {Murg}, \citenamefont {Cirac},\ and\ \citenamefont
  {Verstraete}}]{Pirvu:2010}%
  \BibitemOpen
  \bibfield  {author} {\bibinfo {author} {\bibfnamefont {B}~\bibnamefont
  {Pirvu}}, \bibinfo {author} {\bibfnamefont {V}~\bibnamefont {Murg}}, \bibinfo
  {author} {\bibfnamefont {J~I}\ \bibnamefont {Cirac}}, \ and\ \bibinfo
  {author} {\bibfnamefont {F}~\bibnamefont {Verstraete}},\ }\bibfield  {title}
  {\enquote {\bibinfo {title} {Matrix product operator representations},}\
  }\href {http://stacks.iop.org/1367-2630/12/i=2/a=025012} {\bibfield
  {journal} {\bibinfo  {journal} {New Journal of Physics}\ }\textbf {\bibinfo
  {volume} {12}},\ \bibinfo {pages} {025012} (\bibinfo {year}
  {2010})}\BibitemShut {NoStop}%
\bibitem [{\citenamefont {Lieb}\ and\ \citenamefont
  {Robinson}(1972)}]{Lieb:1972}%
  \BibitemOpen
  \bibfield  {author} {\bibinfo {author} {\bibfnamefont {E.~H.}\ \bibnamefont
  {Lieb}}\ and\ \bibinfo {author} {\bibfnamefont {D.~W.}\ \bibnamefont
  {Robinson}},\ }\href@noop {} {\bibfield  {journal} {\bibinfo  {journal}
  {Commun. Math. Phys.}\ }\textbf {\bibinfo {volume} {28}},\ \bibinfo {pages}
  {251} (\bibinfo {year} {1972})}\BibitemShut {NoStop}%
\bibitem [{\citenamefont {Bravyi}\ \emph {et~al.}(2006)\citenamefont {Bravyi},
  \citenamefont {Hastings},\ and\ \citenamefont {Verstraete}}]{Bravyi:2006}%
  \BibitemOpen
  \bibfield  {author} {\bibinfo {author} {\bibfnamefont {S.}~\bibnamefont
  {Bravyi}}, \bibinfo {author} {\bibfnamefont {M.~B.}\ \bibnamefont
  {Hastings}}, \ and\ \bibinfo {author} {\bibfnamefont {F.}~\bibnamefont
  {Verstraete}},\ }\bibfield  {title} {\enquote {\bibinfo {title}
  {Lieb-robinson bounds and the generation of correlations and topological
  quantum order},}\ }\href {\doibase 10.1103/PhysRevLett.97.050401} {\bibfield
  {journal} {\bibinfo  {journal} {Phys. Rev. Lett.}\ }\textbf {\bibinfo
  {volume} {97}},\ \bibinfo {pages} {050401} (\bibinfo {year}
  {2006})}\BibitemShut {NoStop}%
\bibitem [{\citenamefont {D'Alessio}\ and\ \citenamefont
  {Rigol}(2014)}]{Dalessio:2014}%
  \BibitemOpen
  \bibfield  {author} {\bibinfo {author} {\bibfnamefont {Luca}\ \bibnamefont
  {D'Alessio}}\ and\ \bibinfo {author} {\bibfnamefont {Marcos}\ \bibnamefont
  {Rigol}},\ }\bibfield  {title} {\enquote {\bibinfo {title} {Long-time
  behavior of isolated periodically driven interacting lattice systems},}\
  }\href {\doibase 10.1103/PhysRevX.4.041048} {\bibfield  {journal} {\bibinfo
  {journal} {Phys. Rev. X}\ }\textbf {\bibinfo {volume} {4}},\ \bibinfo {pages}
  {041048} (\bibinfo {year} {2014})}\BibitemShut {NoStop}%
\bibitem [{\citenamefont {Lazarides}\ \emph {et~al.}(2014)\citenamefont
  {Lazarides}, \citenamefont {Das},\ and\ \citenamefont
  {Moessner}}]{Lazarides:2014}%
  \BibitemOpen
  \bibfield  {author} {\bibinfo {author} {\bibfnamefont {Achilleas}\
  \bibnamefont {Lazarides}}, \bibinfo {author} {\bibfnamefont {Arnab}\
  \bibnamefont {Das}}, \ and\ \bibinfo {author} {\bibfnamefont {Roderich}\
  \bibnamefont {Moessner}},\ }\bibfield  {title} {\enquote {\bibinfo {title}
  {Equilibrium states of generic quantum systems subject to periodic
  driving},}\ }\href {\doibase 10.1103/PhysRevE.90.012110} {\bibfield
  {journal} {\bibinfo  {journal} {Phys. Rev. E}\ }\textbf {\bibinfo {volume}
  {90}},\ \bibinfo {pages} {012110} (\bibinfo {year} {2014})}\BibitemShut
  {NoStop}%
\bibitem [{\citenamefont {Ponte}\ \emph {et~al.}(2015)\citenamefont {Ponte},
  \citenamefont {Chandran}, \citenamefont {Papi{\'c}},\ and\ \citenamefont
  {Abanin}}]{Ponte:2014}%
  \BibitemOpen
  \bibfield  {author} {\bibinfo {author} {\bibfnamefont {Pedro}\ \bibnamefont
  {Ponte}}, \bibinfo {author} {\bibfnamefont {Anushya}\ \bibnamefont
  {Chandran}}, \bibinfo {author} {\bibfnamefont {Z.}~\bibnamefont {Papi{\'c}}},
  \ and\ \bibinfo {author} {\bibfnamefont {Dmitry~A.}\ \bibnamefont {Abanin}},\
  }\bibfield  {title} {\enquote {\bibinfo {title} {Periodically driven ergodic
  and many-body localized quantum systems},}\ }\href {\doibase
  10.1016/j.aop.2014.11.008} {\bibfield  {journal} {\bibinfo  {journal} {Annals
  of Physics}\ }\textbf {\bibinfo {volume} {353}},\ \bibinfo {pages} {196}
  (\bibinfo {year} {2015})}\BibitemShut {NoStop}%
\bibitem [{\citenamefont {Prosen}(2002)}]{Prosen:2002}%
  \BibitemOpen
  \bibfield  {author} {\bibinfo {author} {\bibfnamefont {T.}~\bibnamefont
  {Prosen}},\ }\bibfield  {title} {\enquote {\bibinfo {title} {Ruelle
  resonances in quantum many-body dynamics},}\ }\href {\doibase
  10.1088/0305-4470/35/48/102} {\bibfield  {journal} {\bibinfo  {journal} {J.
  Phys. A:Math. Gen}\ }\textbf {\bibinfo {volume} {35}},\ \bibinfo {pages}
  {L737} (\bibinfo {year} {2002})}\BibitemShut {NoStop}%
\bibitem [{\citenamefont {De~Roeck}\ and\ \citenamefont
  {Huveneers}(2014)}]{Roeck:2014}%
  \BibitemOpen
  \bibfield  {author} {\bibinfo {author} {\bibfnamefont {Wojciech}\
  \bibnamefont {De~Roeck}}\ and\ \bibinfo {author} {\bibfnamefont
  {F.}~\bibnamefont {Huveneers}},\ }\bibfield  {title} {\enquote {\bibinfo
  {title} {Scenario for delocalization in translation-invariant systems},}\
  }\href {\doibase 10.1103/PhysRevB.90.165137} {\bibfield  {journal} {\bibinfo
  {journal} {Phys. Rev. B}\ }\textbf {\bibinfo {volume} {90}},\ \bibinfo
  {pages} {165137} (\bibinfo {year} {2014})}\BibitemShut {NoStop}%
\bibitem [{\citenamefont {Yao}\ \emph {et~al.}(2014)\citenamefont {Yao},
  \citenamefont {Laumann}, \citenamefont {Cirac}, \citenamefont {Lukin},\ and\
  \citenamefont {Moore}}]{Yao:2014}%
  \BibitemOpen
  \bibfield  {author} {\bibinfo {author} {\bibfnamefont {N.Y.}\ \bibnamefont
  {Yao}}, \bibinfo {author} {\bibfnamefont {C.~R.}\ \bibnamefont {Laumann}},
  \bibinfo {author} {\bibfnamefont {J.~I.}\ \bibnamefont {Cirac}}, \bibinfo
  {author} {\bibfnamefont {M.D.}\ \bibnamefont {Lukin}}, \ and\ \bibinfo
  {author} {\bibfnamefont {J.E.}\ \bibnamefont {Moore}},\ }\bibfield  {title}
  {\enquote {\bibinfo {title} {{Quasi Many-body Localization in Translation
  Invariant Systems}},}\ }\href@noop {} {\bibfield  {journal} {\bibinfo
  {journal} {ArXiv e-prints}\ } (\bibinfo {year} {2014})},\ \Eprint
  {http://arxiv.org/abs/1410.7407v1} {arXiv:1410.7407v1 [cond-mat.dis-nn]}
  \BibitemShut {NoStop}%
\end{thebibliography}%

\end{document}